\documentclass[3p,times,procedia]{elsarticle}

\usepackage{lineno}

\usepackage{hyperref}
\usepackage{booktabs}
\usepackage{amsmath}
\usepackage{amssymb}
\usepackage{paralist}
\usepackage{multirow}
\usepackage{xspace}
\usepackage{color}
\usepackage{xcolor}
\usepackage{ifthen}
\usepackage{url}
\usepackage{fancybox}
\usepackage{enumitem}
\usepackage{listings}
\usepackage{balance}
\usepackage{ifthen}
\usepackage{subfig}
\usepackage{graphicx}
\usepackage{flushend}
\usepackage{tablefootnote}
\usepackage{colortbl}
\usepackage{pdflscape}

\newenvironment{hassanbox}%
{\begin{center}\vspace{0mm}\noindent\begin{Sbox}\begin{minipage}{0.95\columnwidth}}%
{\end{minipage}\end{Sbox}\fbox{\TheSbox}\end{center}\vspace{0mm}}

\graphicspath{{pics/}}
\DeclareGraphicsExtensions{.pdf,.jpeg,.png}% correct b
\modulolinenumbers[5]

\begin{document}

\newcommand{\eg}{e.g.,}
\newcommand{\ie}{i.e.,}

\newcommand{\activeAPI} {client-used API classes}
\newcommand{\inactiveAPI}{non client-used API classes}

\newcommand{\activeAPIAB} {clientUse}
\newcommand{\inactiveAPIAB}{non clientUse}

\newcommand{\nonapi}{non API classes}

\newcommand{\BC} {breaking~classes}
\newcommand{\nBC} {non-breaking~classes}
\newcommand{\UC} {\textit{nChange}}
\newcommand{\CC} {\textit{change}}

\newcommand{\RC} {\textit{Ref~}}
\newcommand{\RO} {\textit{R$_{opt}$}}

\newcommand{\RqAlpha}{{Using automated tools detection, what is the relationship between API breakages and refactoring activities?}\xspace}

\newcommand{\RqOne}{{To what extent are library maintainers breaking client-used APIs over time?}\xspace}

\newcommand{\RqTwo}{{To what extent are refactoring activities breaking client-used APIs?}\xspace}

\newcommand{\RqTwoTwo}{{What non-refactoring-related code changes are breaking client-used APIs?}\xspace}

\newcommand{\RqThree}{{What refactoring-related code changes are breaking client-used APIs? }\xspace}

\begin{frontmatter}

\title{An Empirical Study on the Impact of Refactoring Activities \\ on Evolving Client-Used APIs}

\author[mymainaddress,mainaddress]{Raula Gaikovina Kula\corref{mycorrespondingauthor}}
\ead{raula-k@is.naist.jp}
\author[mymainaddress,meaddy]{Ali Ouni}
\ead{ali@ist.osaka-u.ac.jp}
\author[mysecondaryaddress]{Daniel M. German}
\ead{dmg@uvic.ca} 
\author[mymainaddress]{Katsuro Inoue}
\ead{inoue@ist.osaka-u.ac.jp}

\address[mymainaddress]{Osaka University, Japan}
\address[mainaddress]{Nara Institute of Science and Technology, Japan}
\address[meaddy]{Ecole de Technologie Superieure Montreal, Canada}
\address[mysecondaryaddress]{University of Victoria, Canada}

\begin{abstract}
\textit{Context: }Refactoring is recognized as an effective practice to maintain evolving software systems. 
For software libraries, we study how library developers refactor their Application Programming Interfaces (APIs), especially when it impacts client users by breaking an API of the library.

\noindent
\textit{Objective:} Our work aims to understand how clients that use a library API are affected by refactoring activities.
We target popular libraries that potentially  impact more library client users.

\noindent
\textit{Method:} We distinguish between library APIs based on their client-usage (refereed to as \textit{client-used APIs}) in order to understand the extent to which API breakages relate to refactorings.
Our tool-based approach allows for a large-scale study across eight libraries (\ie~totaling 183 consecutive versions) with around 900 clients projects. 

\noindent
\textit{Results:}
We find that library maintainers are less likely to break \activeAPI. 
Quantitatively, we find that refactoring activities break less than 37\% of all client-used APIs.
In a more qualitative analysis,  we show two documented cases of where non-refactoring API breaking changes are motivated by other maintenance issues (i.e., bug fix and new features) and involve more complex refactoring operations.

\noindent
\textit{Conclusion:}
Using our automated approach, we find that library developers are less likely to break APIs and tend to break client-used APIs when addressing these maintenance issues.
\end{abstract}


\end{frontmatter}

%\linenumbers

\section{Introduction}\label{sec:introduction}
%context

Software libraries are constantly evolving, either responding to client needs, patching bug fixes or addressing other maintainability concerns.
Refactoring is a controlled and widely-used technique for improving the design of an existing software, especially with modern and large-scale software systems that depend on a large number of third-party libraries.
Fowler recommends refactoring to improve software readability and reusability, while increasing the speed at which developers can write and maintain their code base \cite{Fowler1999,opdyke1992refactoring}. 

The Application Programming Interface (API) are specifications that govern interoperability between a client application and a library. \textit{External APIs} refer to the APIs available for  client usage. 
Since clients solely rely on APIs for `blackbox' access to the library's functionality, \textit{API backward compatibility} is an important consideration for both client and library developers. 
Clients migrating to a newer library version would be particularly concerned with whether previously invoked external APIs in an older version will continue to be invoked without error.
This is known as preserving API compatibility\footnote{ Java standards documentation at \url{http://docs.oracle.com/javase/specs/jls/se8/html/jls-13.html}}. 
Hence,  any API change between two library versions that violates this linkage is known as an \textit{API breakage}.
From a library viewpoint, a developer refactoring an external APIs may not consider the effect its has in affecting a client's chances of adopting the latest version.
Conversely, negligence to refactor the code base may increase the complexity and maintainability efforts (Lehman's 2$^{nd}$ law), leading up to an eventual degradation in software quality (Lehman's 7$^{th}$ law) \cite{Lehman:1996}.

In this work, we conduct an empirical study to explore the relationship between API refactorings and breakages based on actual API usage by clients. 
We distinguish between library APIs based on their client-usage (refereed to as \textit{client-used APIs}) in order to get a deeper understanding on the extent to which API breakages can be related to refactoring activities.
Our investigation covers over 9,700 breaking classes and around 12,900 refactoring operations from eight popular Java libraries, with each library having around 10$\sim$38 consecutive releases.
We observe the following: 
(i) library maintainers are less likely to break client-used APIs compared to other classes of the library, 
(ii) detected refactoring operations only breaking less than 37\% of client-used APIs, qualitatively finding that the (iii) rest (63\%) API breakages are motivated by maintenance issues that are likely to involve more complex refactorings.
Finally, we find that (iv) simple refactorings (\ie~\texttt{move\_method}, \texttt{rename\_method}, \texttt{move\_field}) were less frequently applied to client-used API classes compared to other classes. 

Our main contributions of this paper are three-fold and can be summarized as follows: 
(1) our study involves the investigation of APIs that are used by actual client,
(2) using automated tooling, we conducted a large scale empirical study to investigate API breakages and refactorings and
(3) we present a large dataset of API breakages and refactorings which is publicly available as a replication package at: \url{http://sel.ist.osaka-u.ac.jp/people/raula-k/APIBreakage/}

The rest of the paper is organized as follows. Section \ref{sec:background}
describes the background and definitions. 
Section \ref{sec:method} presents our approach we use in the empirical study. 
Section \ref{sec:rq} details the research questions and what method is used in the study.
We then show our results in Section \ref{sec:results}, with discussion of implications and threats of the study in Section \ref{sec:discussion}. 
Section \ref{sec:related} surveys related work.
Finally, Section \ref{sec:conclusion} concludes the paper and presents future research directions.

\section{Basic Concepts \& Definitions}\label{sec:background}

This section provides the necessary background and concepts that are prerequisites to understand the conducted study.
%context
\subsection{Backward Compatibility of APIs}

The precise definition of backward compatibility depends in part on the Java language's notion of binary compatibility\footnote{documentation at \url{http://docs.oracle.com/javase/specs/jls/se8/html/jls-13.html}}: 
\begin{quote}
	\textit{ ``binary compatible with (equivalently, does not break binary compatibility with)  pre-existing binaries if pre-existing binaries that previously linked without error will continue to link without error."}
\end{quote}
Importantly, a class or interface should treat its accessible members (method and fields) and constructors, their existence and behavior, as a \textit{contract} with its users. 

In this paper, we define that any changes violating this contract are said to cause an API breakage between the library and its client user.
We show two examples of API breakages.
The first example of an API breakage is when a method name is modified (i.e., renamed or deleted method).  
For instance, the removal of the method in a class could break the API linkage, resulting in a \texttt{NoClassDefFound} exception error to the client application.
Conversely, adding parameters (\ie~adding new fields, methods, or constructors) to an existing class or interface usually does not break an API.

The second example of an API breakage is when third-parties cause an API breakage to the library, which then indirectly breaks the client.
In many cases, a library is also a client user of other libraries within their environment.
For instance, any changes to the library's environment such as an update to the Java Development Kit (JDK) may break a method in the library, and therefore ripples its effect to any client user of this library API. 

%
%As a final suggestion, although I could understand the Examples in Figure 9 after some time, I recommend the authors to add a brief example of this type of problem at the end of Section 2.1 as they did with other types of problems (change in the method signature). You could add a sentence explaining the case where the method uses some code that depends on a particular version of the JVM or other similar example.

\subsection{Refactoring Activities and API breakages}
Refactoring is a disciplined engineering practice that restructures an existing code by altering its internal structure without changing its external behavior \cite{Fowler1999}.
Fowler discusses around seventy various refactorings, which can be either simple or become quite complex.
In this paper, we determine if any of the API breakages is related to a refactoring activity.
Formally, we define a Refactoring Operation (\textit{\RO}) as an atomic refactoring change applied between two library versions. 

\subsection{API Categorization Based on Client Usage}
In this paper, we are interested in the APIs actually used by a client application, assuming that a code change between a client-used API will cause a breakage to that contract between library and client user.
To investigate the extent of which developers are breaking their APIs, we must first define the usage dimension of an API. 
In reality, not all public entities (APIs) are intended for client usage.
Based on a developer's intended use, an API of a library can either be \texttt{external} or \texttt{internal}.

\begin{itemize}
	\item \textbf{External APIs} - are APIs designed by library maintainers for usage by clients.
	
	\item \textbf{Internal APIs} - are APIs intended only for internal usage by the library code itself. 
	
\end{itemize}

An internal API may exist for several reasons. For instance, the \texttt{Finalizer} class within \texttt{base.internal} package of the \texttt{google-guava}$_{17.0}$ documented \cite{caseExample}:

\begin{quote}
	\emph{While this class is public, we consider it to be *internal* and not part of our published API. It is public so we can access it reflectively across class loaders in secure environments}.
\end{quote}
In an ideal world, internal APIs are never used by any client. However, in reality internal APIs may be subjected to client usage. For instance, Businge \textit{et al.} found that a large proportion of plugins used the Eclipse framework internal APIs \cite{DBLP:journals/sqj/BusingeSB15}. Moreover, concepts such as the Application Binary Interfaces (ABIs) \cite{Forman:1995} and the OSGi framework \cite{Baul2009} have been proposed to differentiate between the two API types. However, unless explicitly documented, it is extremely difficult to distinguish between \textit{external} or \textit{internal} APIs.

\begin{figure}[t]
	\centering
	\includegraphics[width=.6\textwidth]{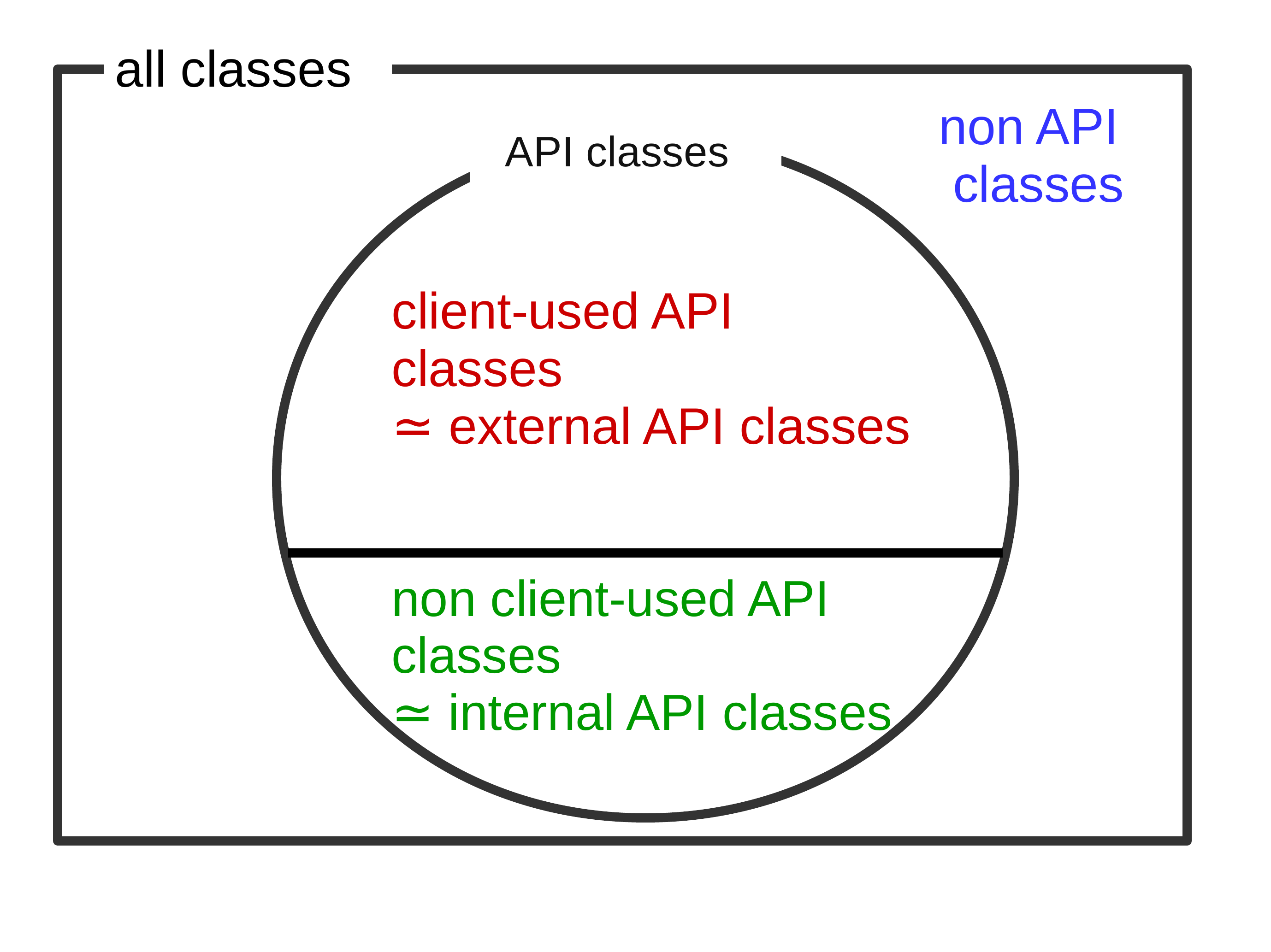}
	\caption{A conceptual composition of all library class types. The venn diagram shows the relationship between (a) client-used API, (b) non client-used API and (c) non API class types.}
	\label{fig:vernn}
\end{figure}

As shown in Figure \ref{fig:vernn}, we describe the different class categories of a library. To distinguish between external and internal APIs, we propose a method to approximate external API classes by mining actual usage by clients, defined as  \textit{\activeAPI}. Details of the method are explained in the subsequent methodology subsection.
All library class categories are defined as follows:

\begin{itemize}
	\item \textbf{API class} -  is a  class that has at least one public entity (\ie~method and field members) and accessible by any client user. 
	%API classes are either external or internal. \sloppypar{\ie~$\textit{API classes}=\textit{internal} ~\cup \textit{external~API~classes}$}.
	
	\item \textbf{non API class} - is a class that contains no API entities, i.e., private or protected.
	%\sloppypar{\ie~$\textit{\nonapi} \neq \textit{API classes}$}.
	
	\item \textbf{client-used API class~(\activeAPIAB)} - is an API class that is used by at least one client. 
	It is an approximation of the external APIs. 
	The set of client-used API classes should ideally cover all external APIs. However, there exist cases when a client uses an internal API. 
	\sloppypar{(\ie~$\textit{\activeAPI}\simeq \textit{external~API~classes}$)}
	
	\item \textbf{non client-used API class~(\inactiveAPIAB)} - is an API class that is not used by any client. The set of all non client-used API classes should cover all internal APIs. 
	\sloppypar{(\ie~$\textit{\inactiveAPI}\simeq internal~API~classes$)}
\end{itemize}

%As for the rest of the library, we term \textit{non APIs} as the rest of private or protected entities.
Henceforth, we classify our API breakages at the class-level.  
Classes are then classified as either:
\begin{itemize}
	\item \textbf{breaking class} - is a changed class that is breaking its API in either class, method or field levels such as rename/move/delete changes.
	\item \textbf{non breaking class} - a changed class that does not affect API compatibility.
\end{itemize}

%In this study, we identify refactorings in relation to \activeAPI. 
We then explore the extent to which breaking changes to client-used API are caused by refactoring activities.
As defined in the Section \ref{sec:background}, \textit{\RO} is an atomic refactoring operation applied between two library versions.
We now introduce the following terminologies related to \textit{\RO}:
\begin{itemize}
	\item \textbf{Ref class} - is a changed API class where at least one \RO~has been applied to any of its elements. (\ie~field, methods or class attributes).
	
	\item \textbf{\RO~density} - refers to the number of \RO~applied per class.
\end{itemize}

\section{Approach}\label{sec:method}
In this section, we first present our case study libraries and methodology used in the empirical study. 
Our method includes (1) categorization of API based on client usage, (2) API breakage detection and (3) API refactoring detection.  

%First we describe how we mine \activeAPI~and detect refactorings. 
%Then, we detail the research method to address each research question.
\begin{table*}[b]
	\fontsize{10}{10}\selectfont
	\centering
	\caption{Studied libraries showing the releases range, number of versions, time period, and the range of number of classes per library  (min-max).}
	\label{tab:targetLib}
	\begin{tabular}{@{}lrccrc@{}}
		\toprule
		\multicolumn{1}{l}{Library} & \multicolumn{1}{c}{Release range} & \multicolumn{1}{c}{\#Versions}  &\multicolumn{1}{c}{Releases Time Period}&\multicolumn{1}{c}{ \# Classes (min $\sim$ max)}\\\midrule
		\textsc{guava}& r03 $\sim$ 18.0 & 22 & Apr 10 $\sim$ Aug 14& 727 $\sim$ 1763\\
		\textsc{httpclient} & 4.0 $\sim$ 4.5 & 25 & Aug 09 $\sim$ May 15 & 230 $\sim$ 460\\
		\textsc{javassist} & 2.5.1  $\sim$ 3.19.0 & 28 & Feb 06$\sim$ Jan 15 & 187 $\sim$ 334\\ 
		\textsc{jdom} & 1.1 $\sim$ 2.0.6 & 10 & Sept 04 $\sim$ Feb 15 & 73 $\sim$ 258\\ 
		\textsc{joda-time} & 0.95 $\sim$ 2.8 & 22 & Nov 05 $\sim$ May 15 & 191 $\sim$ 246\\ 
		\textsc{log4j} &1.1.3 $\sim$ 1.2.17 & 17 & Jun 01 $\sim$ May 12 &242 $\sim$ 974 \\
		\textsc{slf4j} &1.1.0 $\sim$ 1.7.12 & 38 & Dec 06 $\sim$ Mar 15& 11 $\sim$ 28\\
		\textsc{xerces}  &1.2.3 $\sim$ 2.11.0 & 21 & Dec 00 $\sim$ Nov 10 &580 $\sim$ 1652\\
		
		\bottomrule
	\end{tabular}
\end{table*}

\subsection{Subject Libraries}
We used a systematic method to select our subject libraries. 
Our selection of these libraries is based on the following criteria: (1) have a large enough client-user API usage and (2) have sufficient evolution history. 
Additionally, we required diverse libraries that (3) are from different application domains and (4) have been extensively studied in related work. 
This criteria was used to select libraries from a set of 2,500 client projects collected from GitHub.

Table \ref{tab:targetLib} shows all 183 library versions from the eight selected libraries. 
For each library, we collected 10 to 38 different library versions. 
All libraries constitute a large client-base and are from different application domains. 
Moreover, three out of the eight subject libraries were used in prior work \cite{Cossette2012,Dig2006,Kapur2010}. 
The chosen studied libraries range from being testing, logging, utilities and web-based libraries. 
As shown in the table, we selected \textsc{guava} \cite{guavaURL}, \textsc{httpclient} \cite{httpclientURL}, \textsc{javassist} \cite{javassistURL}, \textsc{jdom} \cite{jdomURL}, \textsc{joda-time} \cite{jodatimeURL}, \textsc{log4j} \cite{log4jURL}, 
\textsc{slf4j} \cite{slf4jURL} and \textsc{xerces} \cite{xercesURL}
%is  an abstraction of logging framworks.
% is a library used for http web protocols, while is a logger.
%Finally  %\footnote{We include official binaries named Jdom and Jdom2} manuscript
% and  %\footnote{We include offically released xerces andmanuscript xercesImpl} 
% are XML parser libraries.
For all libraries, we only selected consecutive version releases, ignoring release candidates. 
Only the official binaries and available source code for each library were used in this study. 

\begin{figure*}
	\centering
	\subfloat[\textsc{guava}]{\label{fig:figGuava}%
		\includegraphics[width=0.4\columnwidth]{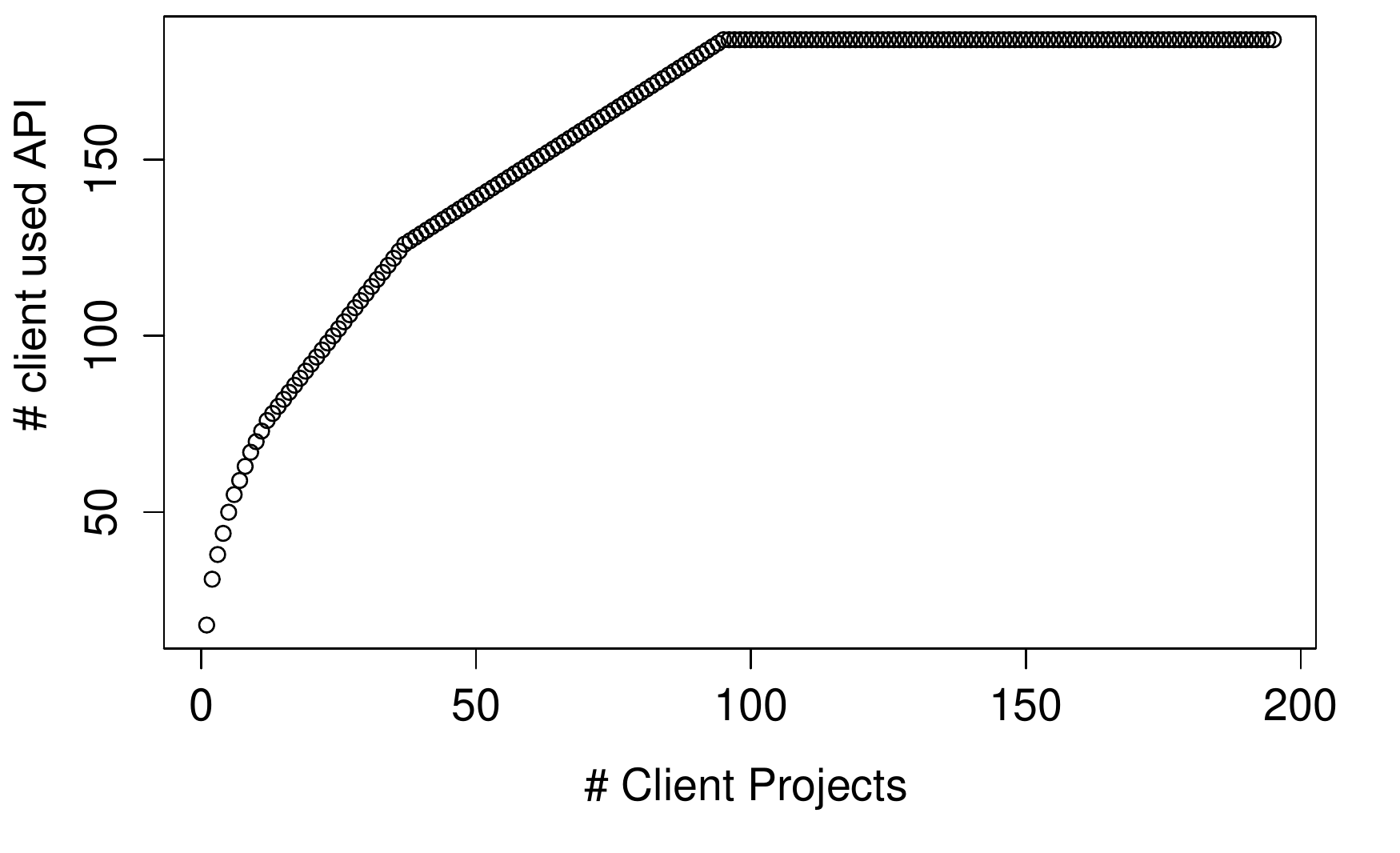}
	}
	\subfloat[\textsc{httpclient}]{\label{fig:activehttpclient}
		\includegraphics[width=0.4\columnwidth]{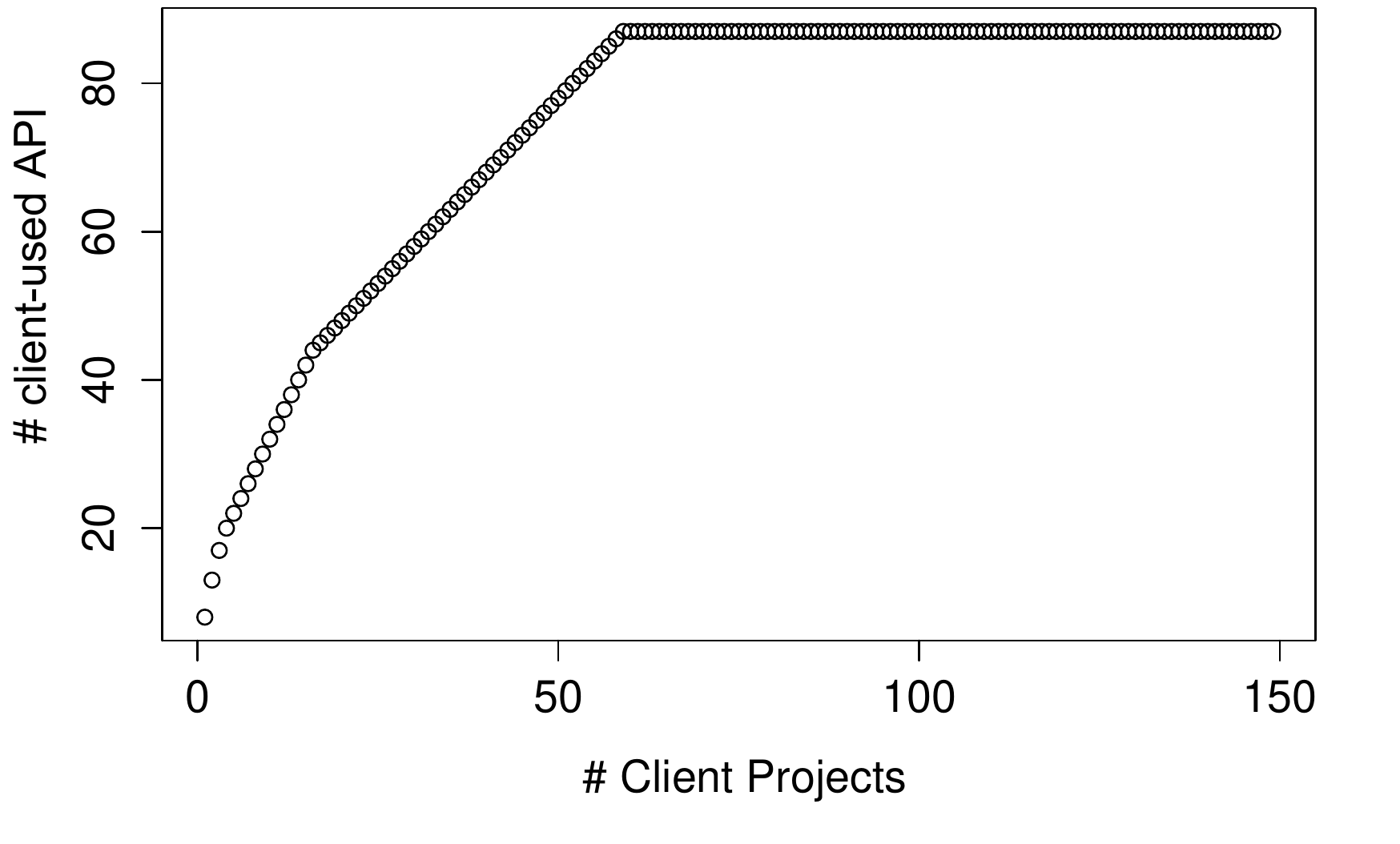}
	}\hfill
	\subfloat[\textsc{javassist}]	{\label{fig:activeJavassist}
		\includegraphics[width=0.4\columnwidth]{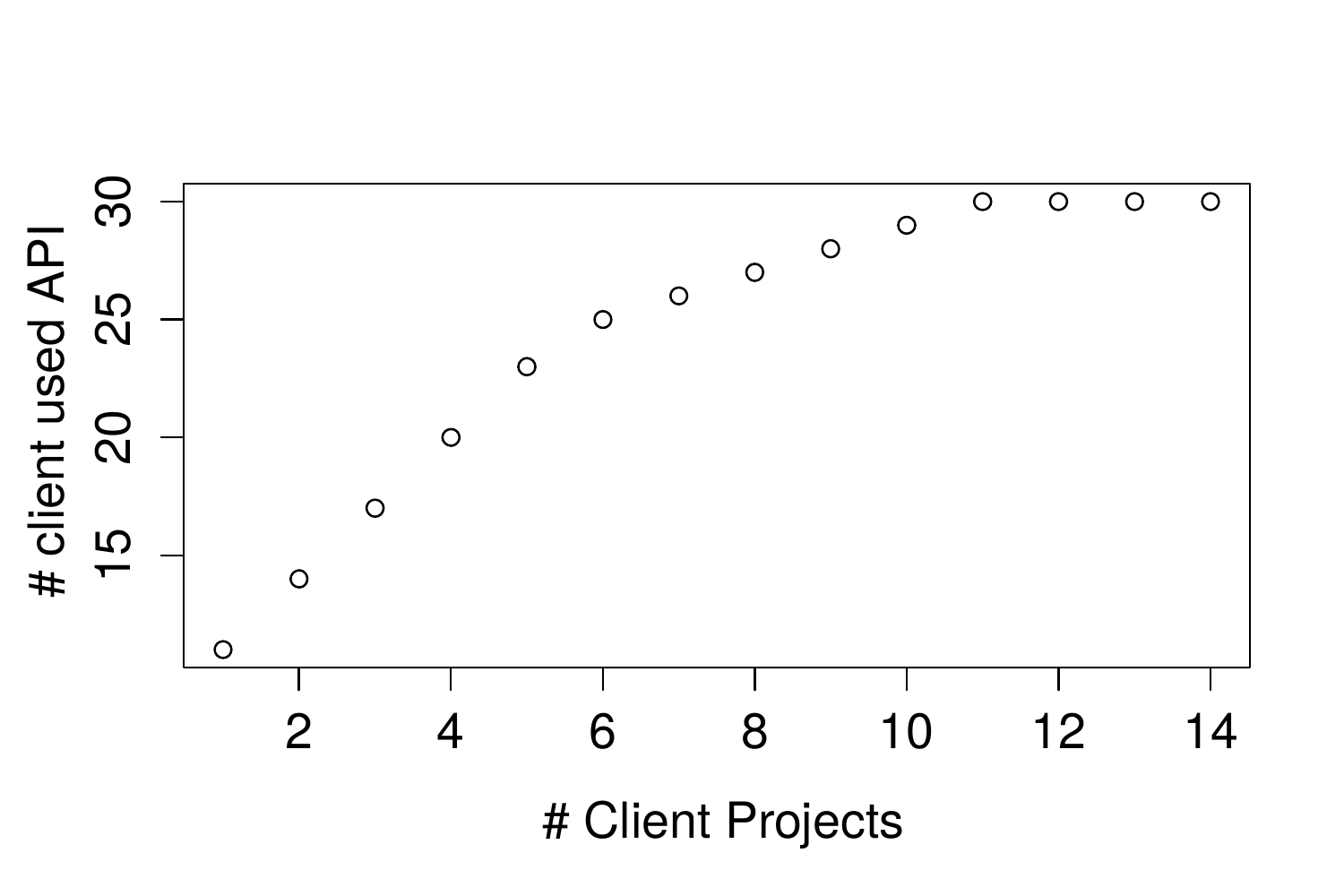}
	}	
	\subfloat[\textsc{jdom}]{\label{fig:activeJdom}
		\includegraphics[width=0.4\columnwidth]{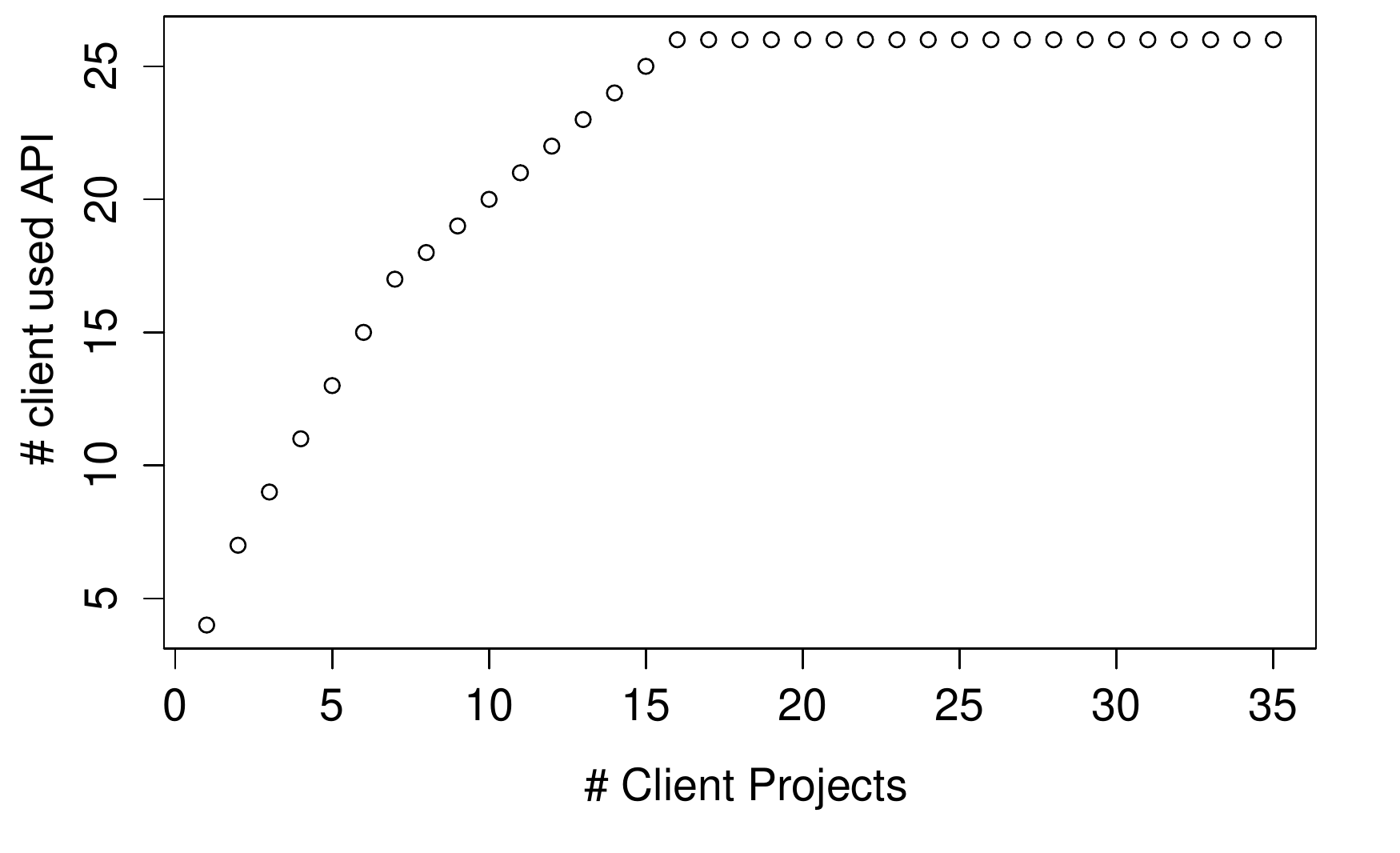}
	}\hfill
	\subfloat[\textsc{Joda-time}]{\label{fig:activeJoda}
		\includegraphics[width=0.4\columnwidth]{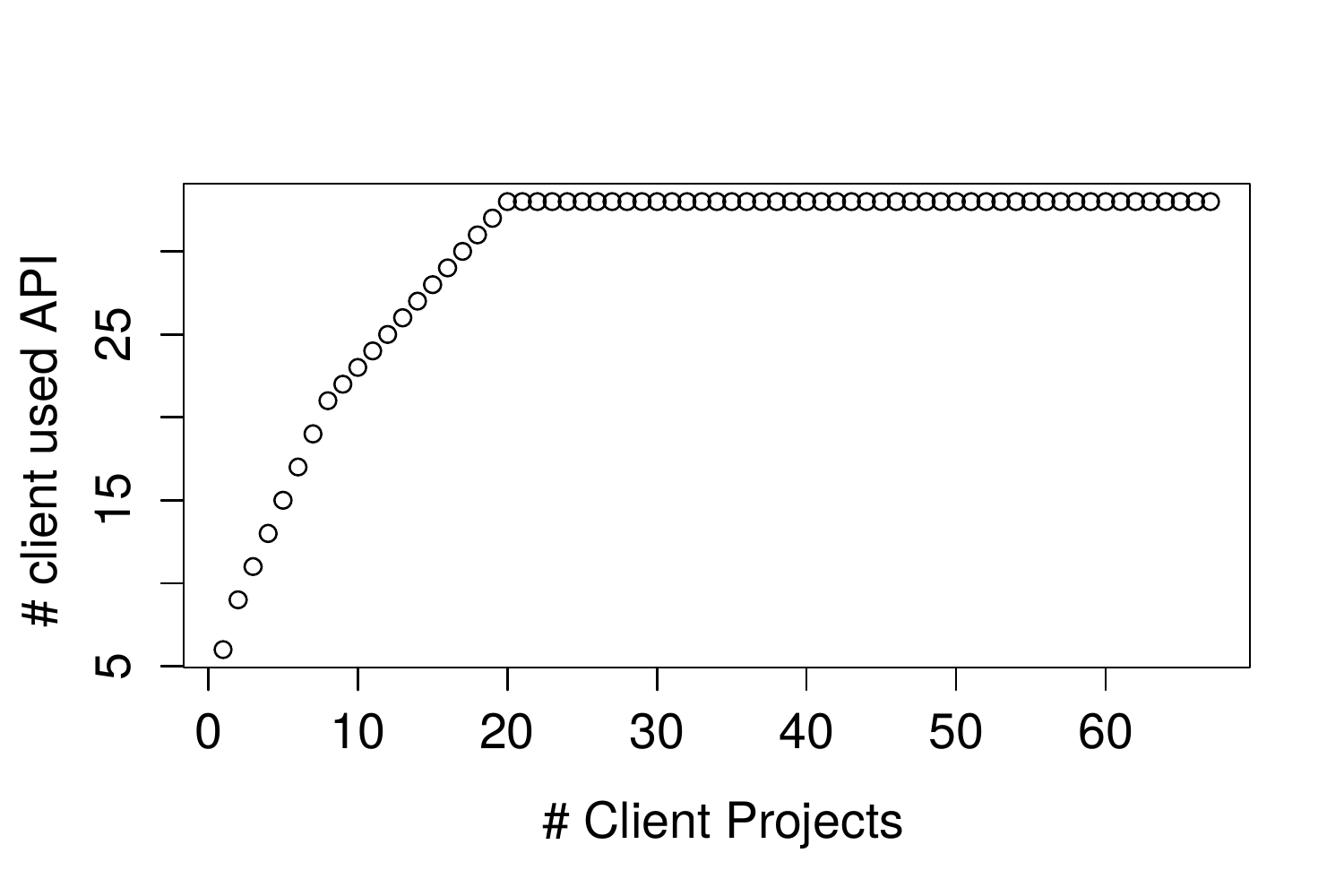}
	}
	\subfloat[\textsc{log4j}]{\label{fig:activeLog4j}
		\includegraphics[width=0.4\columnwidth]{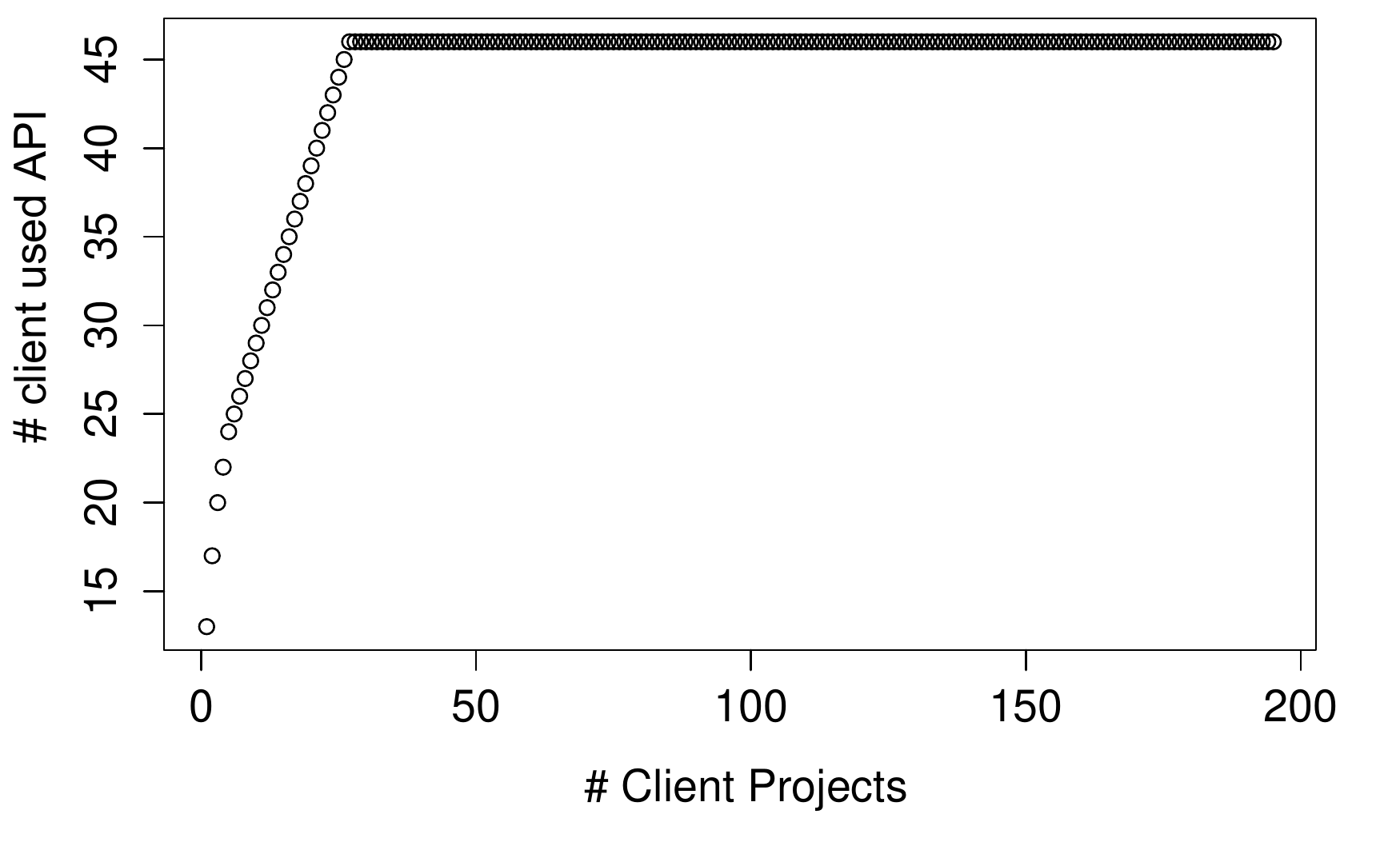}
	}\hfill
	\subfloat[\textsc{slf4j}]	{\label{fig:activeslf4j}
		\includegraphics[width=0.4\columnwidth]{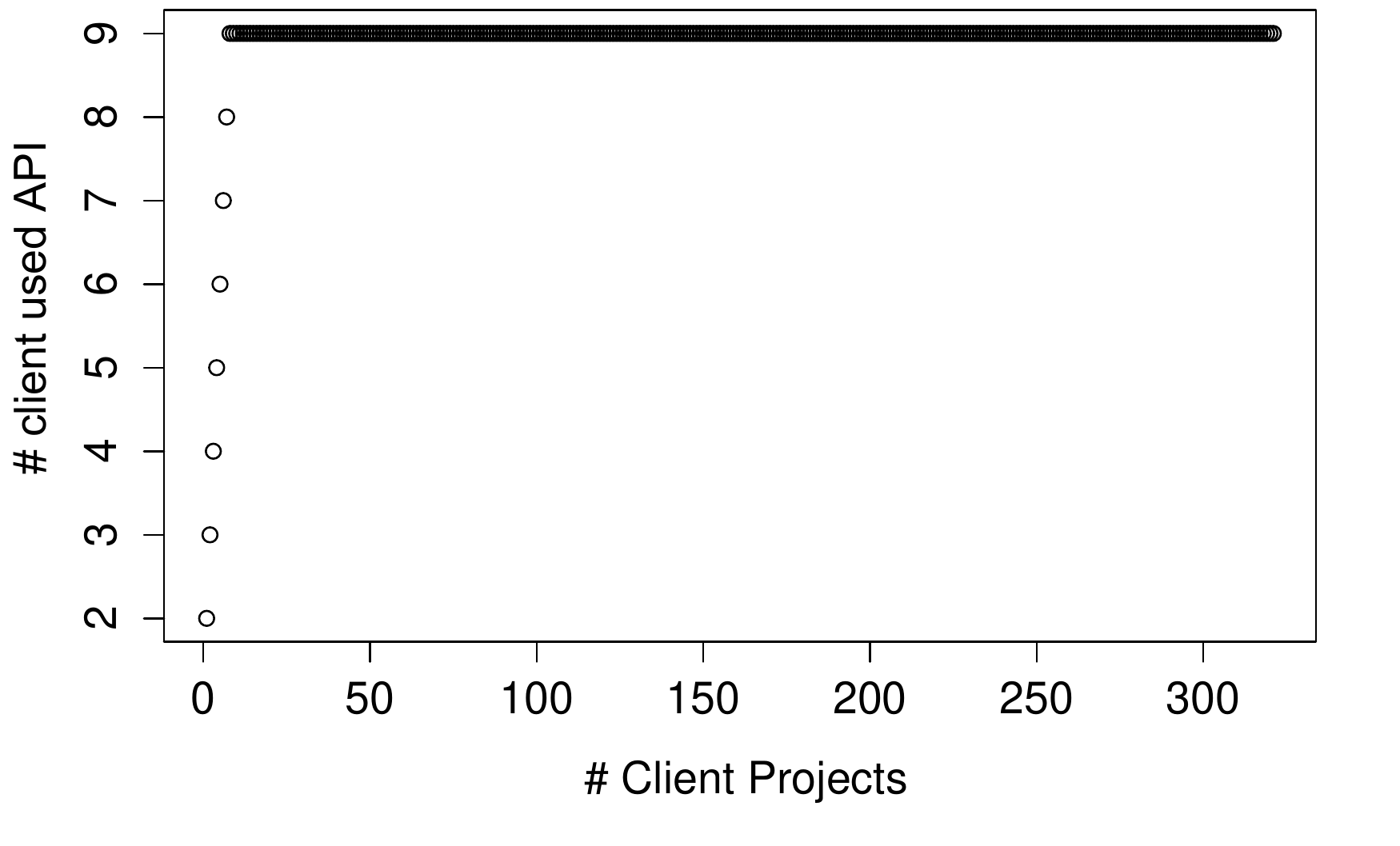}
	}
	\subfloat[\textsc{xerces}]{\label{fig:activeXerces}
		\includegraphics[width=0.4\columnwidth]{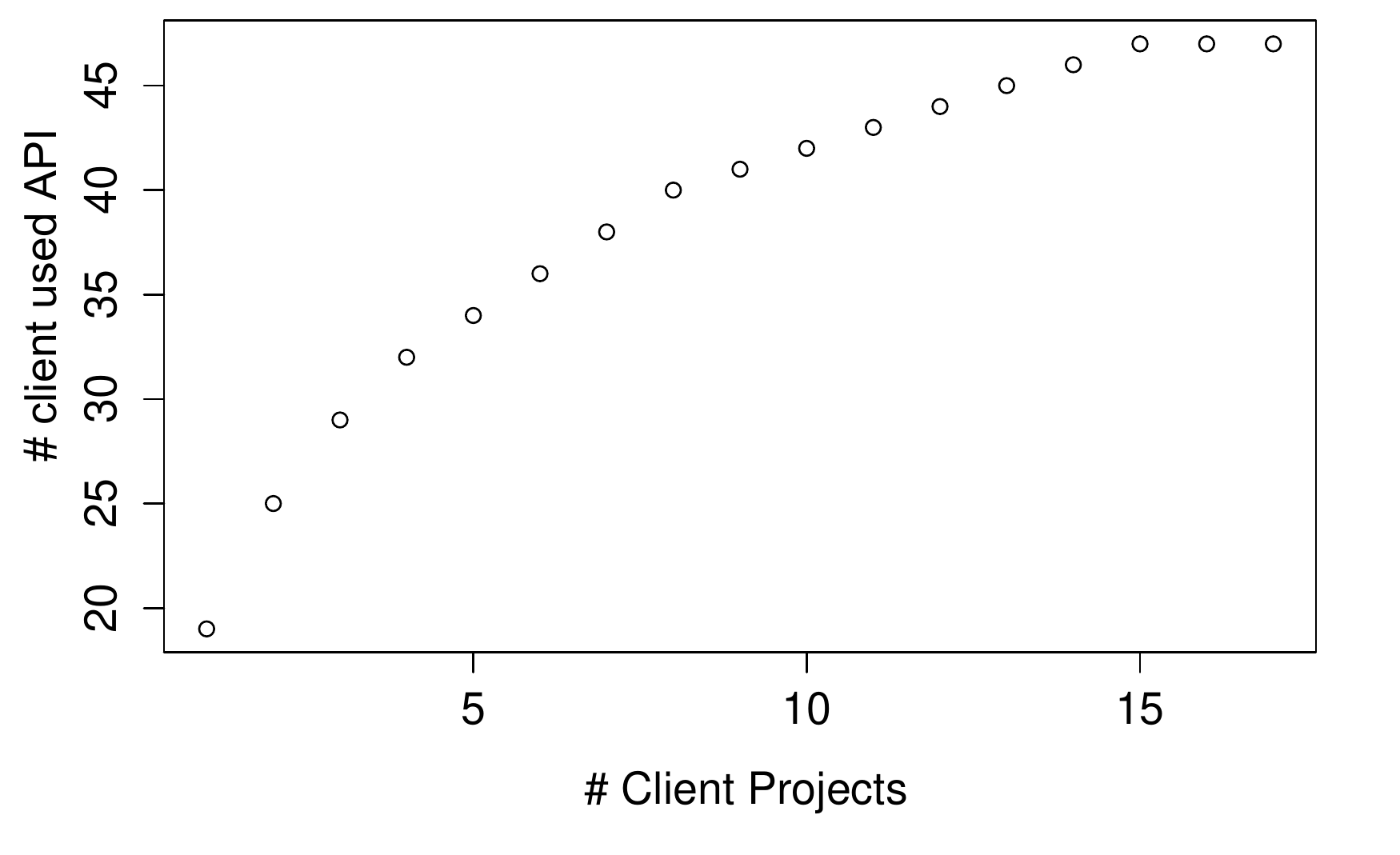}
	}\hfill

	\caption{Cumulative count of \activeAPI~(x-axis) represented as a function over the number of client projects (y-axis). 
		The saturation function (coasting of the curve) indicates that a stable number of \activeAPI~classes have been reached (See Table \ref{tab:minedLib}).}
	\label{fig:saturationPlot}
\end{figure*}

\subsection{Client-Used API Extraction Method}

Actual client usage is needed to distinguish between \textit{external} or \textit{internal} APIs.
Specifically, we would need to compile each individual client system to know what APIs are used by clients.
To enable a large scale analysis, we use the \texttt{jcabi-aether} \cite{jcabiURL} library and \texttt{JavaCompiler}  (\textit{ver.1.6}) Eclipse compiler \cite{javaCompilerURL} to dynamically compile and log all client-loaded classes.
As a result, we are able to extract the fully qualified library class name of all external APIs for many clients.

\begin{table}[t]
	\fontsize{10}{10}\selectfont
	\centering
	\caption{Collected \activeAPI~as shown in Figure \ref{fig:saturationPlot}}.
	\label{tab:minedLib}
	\begin{tabular}{@{}lcccc@{}}
		\toprule
		\multicolumn{1}{c}{} &  & \multicolumn{2}{c}{At Saturation Point (SP)}  \\
		\cmidrule(r){3-4}
		& \multicolumn{1}{c}{\# Collected Clients}   & used clients at SP & \activeAPI~at SP \\ \midrule
		\cellcolor{gray!25}{\textsc{Guava}}  & \cellcolor{gray!25}{195}  & \cellcolor{gray!25}{98} & \cellcolor{gray!25}{184}\\
		\textsc{httpclient}&149  &67& 87\\
		\textsc{Javassist}& 14& 11& 30\\
		\textsc{Jdom}& 35&16& 26\\
		\textsc{Joda-time}& 69& 20& 27\\
		\textsc{log4j}& 195&36& 46\\ 
		\textsc{Slf4j}  & 321  & 20& 9 \\
		\textsc{Xerces}&17 &15& 47 \\
		\bottomrule
		All clients&995&&&\\
	\end{tabular}
\end{table}

One of the main challenges to determine client-used API collection is the coverage of all external APIs.
Hence, our technique consists of continuously collecting client systems until full coverage is reached (i.e., no more APIs are used). 
%We reach a coverage when all client all external APIs as \activeAPI. 
We coin this coverage as the  \textit{saturation point} reached for a library version. 
So instead of trying to compile as many clients are possible, we use the saturation point as a heuristic to show that enough clients have been collected.
Figure \ref{fig:saturationPlot} and Table \ref{tab:minedLib} shows the saturation point for our case studies. 
The saturation point is represented as a cumulative count of \activeAPI~(x-axis) represented as a function over the number of client projects (y-axis), with the coasting of the curve assuring confidence that a stable number of \activeAPI~have been reached. For example, of the 195 collected clients, guava reached a saturation with 98 client systems to cover 184 API classes. 
It is important to note that each project was selected at random, making the formation of the curve unintentional. 
The table also summarizes the number of client GitHub projects that we mined for each of the eight subject libraries (total code base size of 600GB).
To ensure maturity and quality of the client projects, the projects dataset only includes java projects that had at least 100 commits.
We ran experiments for about 30 days.
The process of client-used API collection of a single project took between 10 min $\sim$ 3 hours. 
%We specifically selected projects that included any of the target libraries.

\subsection{API Breakage Detection Method}
\label{sec:researchMethod}
%We rely on automated tools to identify 
In recent times, state--of--the--art API breakage detection tools \cite{japicmp, clirr, jacc, jdiff, revapi}  have been extensively used by both researchers \cite{Jezek2015}, \cite{Raemaekers2014} and practitioners \cite{guavaURL}, \cite{httpclientURL} alike, especially for a systematic comparison of API checking backward incompatibilities between library versions\footnote{For instance, developers of the google guava library, use \texttt{JDiff} to report changes between two versions, e.g.,  API changes from guava v18 to v19 are at \url{http://google.github.io/guava/releases/19.0/api/diffs/}}
As noted by Raemaekers \cite{Raemaekers2014}, these tools are underestimations-- as all detected breaking API changes will definitely break an API but some binary compatible APIs could be semantically incompatible. 

To identify the API differences between two library binaries, we use the \texttt{Japi-cmp} library  \cite{japicmpURL}.
Similar to other tools, \texttt{Japi-cmp} is able to detect and differentiate changes in instrumented and generated classes to determine binary compatibility as well as public or private accessibility. 
Using the definitions in Section \ref{sec:background}, we then map and label all classes as either breaking or non-breaking.
Overall, the resulting dataset consists of over 9,700 detected breaking classes from the eight libraries.

\subsection{Refactorings Detection Method}
To automatically collect \textit{\RO}~applied between the two versions, we use the state--of--the--art   \texttt{Ref-Finder} \cite{prete2010template} tool.
Based on template logic rules, the tool identifies up to different 52 refactoring types between two versions. 
It is important to note that the collected refactorings are structural, only detectable by mechanical transformations; \textit{`` Ref-Finder does not include changes that may either require restricted conditions to be met, or to some degree of additional specification from a developer that could not be automatically inferred by a tool''} \cite{Cossette2012}.   
As a result, our dataset consists of 12,900 \textit{\RO} from all eight libraries.

\subsection{Mapping Refactorings to API Breakages}
\label{sec:breakMap}
 \begin{figure}
 	\centering
 	\includegraphics[width=.6\textwidth]{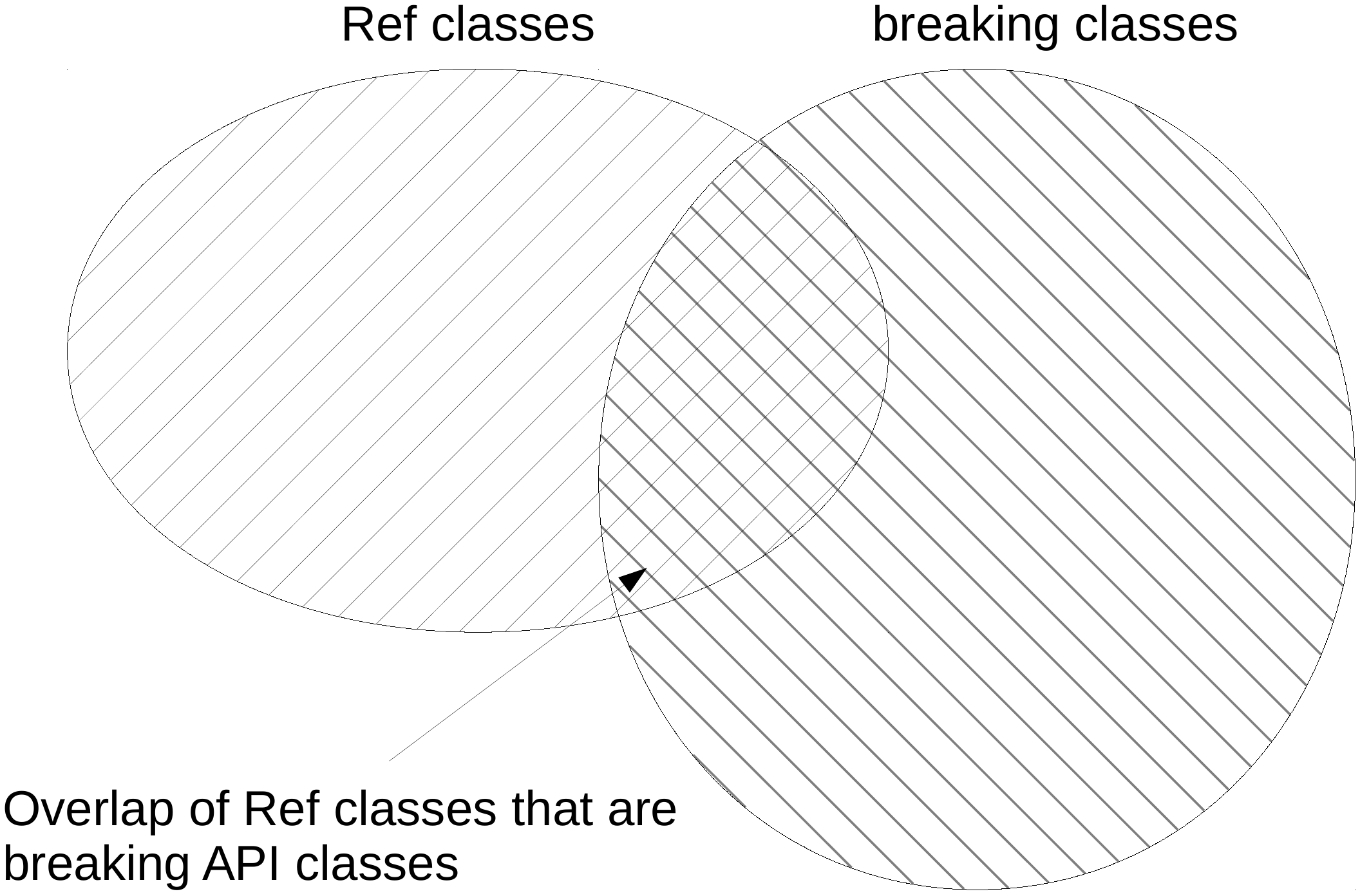}
 	\caption{Venn diagram of the overlapping relations of refactored and breaking classes.}
 	\label{fig:RQ2RefBreak} 
 \end{figure}
 
The study involves a mapping between the collected \texttt{Ref} and \texttt{breaking} classes, where a \texttt{Ref class}  contains at least one \textit{\RO}.
Figure \ref{fig:RQ2RefBreak} describes this mapping as an intersection between \textit{breaking classes} and \textit{Ref classes}. It is important to note false positives, where the tools detect refactorings in unchanged classes.
Upon manual inspection of some cases, we confirmed these were false positives as the classes were unchanged.
As a result, we semi-automatically identified and discarded 2,100 instances of such false positives, finally leaving us with 10,800 \textit{\RO} from all eight libraries.
 
A simple example of a refactoring that breaks API can be seen with the \texttt{com.google.common.collect.ImmutableMultiset} of the \textsc{Guava} library\footnote{ the API change at \url{http://google.github.io/guava/releases/12.0/api/diffs/changes/com.google.common.collect.ImmutableMultiset.html\#methods}}. According to the API Diff report, the \texttt{ImmutableMultiset<E> of(E[])} method (\ie~which takes   \texttt{E[]} and returns an immutable multiset) was removed between version 11.0.02 and 12.0. 
In this example, our approach automatically detects this change as the \texttt{remove\_method} \RO. 
The official Java documentation states that \textit{`deleting a method or constructor from a class  breaks compatibility with any pre-existing binary that referenced this method or constructor; a \texttt{NoSuchMethodError} may be thrown when such a reference from a pre-existing binary is linked. 
Such an error will occur only if no method with a matching signature and return type is declared in a superclass'}.
% please add the example: https://github.com/google/guava/commit/7ebdd5768ad1a5f839a9c78f42606f91bfc22546
% https://github.com/dCache/dcache
 
% for entryset() 
%https://github.com/google/guava/commit/bd3d46ff61aab113bac98311b77c1d5f6ce1334e
%https://codereview.appspot.com/5753064/patch/2003/24

\section{Empirical Study}\label{sec:rq}
In this section, we present the goals and motivation, followed by the method used to address each research question.

\subsection{Research Questions}
\label{sec:motiv}
Our motivation is to inspect the relationship between refactorings and API breakages.
Related, Dig and Johnson \cite{Dig2006} manually inspected library release notes for documented API changes to investigate the role of refactoring during API evolution of a library. 
They cited two reasons why they preferred a manual analysis over the use of automated tools: (1) \textit{`since most API changes follow a long deprecation replace-remove cycle, an obsolete API can coexist with the new API for a long time'} and (2) some behavioral refactoring cases that \textit{`would have been misinterpreted by a tool, but a human expert can easily spot'}.
In this study, we find that state--of--the--art tools are now able to detect deprecations, thus negating the first reason. Additionally, we find that the automated approach is not as reliant on  documentation.
%In fact, the automated approach has additional benefits.
%For instance, it reduces some errors of manual inspection and heuristics.
%In the Dig study, they relied on release notes of developers as a heuristic for API changes, \textit{`starting from the change logs that describe the API changes for each release'}. 
%The advantage of tooling is the ability to syntactically compare and report all applied changes between two library versions.
%Other reasons are that the results are easy to replicate and the tool allows for a more uniformed coverage over a larger range of client projects.   

Our goal in this study is to use an automated approach to investigate how client usage-APIs are affected by the refactoring activities.
The automated approach has the benefit of reducing manual inspection and heuristic errors and enables a large-scale empirical study.
We designed a rigorous quantitative empirical study, formulating the following research questions:

\begin{itemize}
	\item \textit{{(RQ1). \RqOne}} We want to understand the API breaking tendencies of library maintainers. 
%	This results will determine if library maintainers avoid violating the contract with their client users.
	\item \textit{{(RQ2). \RqTwo}} Sometimes API breakages are unavoidable, even for the more popular client-used APIs. Prior work indicates that refactoring is common with API changes. Therefore, we want to understand how much of client-used API breakage is related to refactoring activities.
\end{itemize}

In RQ2, we identified many API breakages not related to refactoring activities.
We then formulated RQ3 and RQ4 for a deeper analysis of the detected changes (both refactoring and non refactoring related) that break client-used APIs:

\begin{itemize}
	\item \textit{{(RQ3). \RqTwoTwo}}
	Specifically, our motivation is to understand what API breaking changes are not related to refactorings.
	
	\item \textit{{(RQ4). \RqThree}} 
	From the perspective of all refactoring activities, we would like to understand (i) how much and (ii) types of refactoring operations that are breaking client-used APIs.

\end{itemize}

\subsection{Research Method for RQ1}
\begin{table}[t]
	\fontsize{10}{10}\selectfont
%	\centering
	\caption{Library Class Categories Incompatibility Matrix}.
	\label{tab:breakTypes}
	\begin{tabular}{lccc}
		\toprule
		\multicolumn{1}{c}{} & \multicolumn{1}{c}{Compatible Change} & \multicolumn{2}{c}{Incompatible Changes ($\textit{break}_{change}(L_v)$)}\\
		\midrule
		\cellcolor{red!25} \textsc{client-used API}&API compatible &\cellcolor{red!25} API Breaking code change\\
		\cellcolor{green!25} \textsc{non client-used API} & API compatible &Incompatible change unintended for client \\
		\cellcolor{cyan!50} \textsc{non API}& Not affect client& Incompatible change does not affect client\\ 
		\bottomrule
	\end{tabular}
\end{table}

To answer RQ1, we followed two steps.
First, we studied consecutive versions of a library to understand the library evolution.
The goal is to study how (i) \activeAPI,  (ii) \inactiveAPI~and (iii) \nonapi~evolve over several consecutive versions.
Next, we investigate the number of code changes that lead to incompatibility with respect to the different class categories that we defined above. 
Since the tool is only able to compare two versions at a time, we performed a side-by-side (\ie~each comparison is the current version against the immediate successive library version).
We introduce a normalized metric namely \textit{break}$_{change}$ to describe the rate of the number of breaking changes over all class changes at that version release as defined in  Equation \ref{eq:M1}:% this breakage rate For each library version $L_v$, let : 

\begin{equation}
\label{eq:M1}
\textit{break}_{change}(L_v) = \frac{|\BC|}{|all~changed~classes| } 
\end{equation}

\noindent where $L_v$ refers to a given library version and ranges from 0 $\leqslant$ ${break}_{change}$ $\leqslant$ 1 for each class category of $L_v$. Values that are closer to 1 indicate that there are more breakages per class changes. 

Table \ref{tab:breakTypes} shows the $\textit{break}_{change}(L_v)$ metric interpretation based on the class type. Hence, the $\textit{break}_{change}(L_v)$ metric has different interpretations based on the class type. For instance, for non API classes, the metric shows significant changes that do not affect clients.
We believe that it is important to track which classes are more prone to incompatible code changes.
To assess the significance of breakages between the different library class categories, we use the Kruskal Wallis and Mann-Whitney non-parametric test. The null hypothesis would state no statistical difference between the class types.
Furthermore, to assess the difference magnitude, we study the effect size based on Cohen's \textit{d} \cite{tagkey1977iii}. The effect size is considered: (1) small if 0.2 $\leqslant$ \textit{d} $<$ 0.5, (2) medium if 0.5 $\leqslant$ \textit{d} $<$ 0.8, or (3) large if \textit{d} $\geqslant$ 0.8. For the effect size, we use the Mann-Whitney tests with Bonferroni correction.

\subsection{Research Method for RQ2}
% \begin{figure}[t]
% 	\centering
% 	\includegraphics[width=.6\textwidth]{refBreak-cropped}
% 	\caption{Venn diagram of the \textit{breaking--to--Ref} and \textit{Ref--to--breaking} rates. Both metrics are ratios of the overlapping relations of the classes with respect to the totals of breaking classes (\textit{Ref--to--breaking rate}) and Ref classes (\textit{breaking--to--Ref rate}).}
% 	\label{fig:RQ2RefBreak} 
% \end{figure}

For RQ2, our method is to identify library refactorings that are applied to \activeAPI. 
We followed two steps. To analyze the impact of the refactoring activities, we first identified for each library the (i) number of Ref classes and (ii) \RO~density. %The \RO~density refers to the number of \RO~applied per class category. 
We then identified the \texttt{Ref classes} that are breaking.
%As shown in Figure \ref{fig:RQ2RefBreak}, we describe these as the intersection between \textit{breaking classes} and \textit{Ref classes}.
To map refactorings to API breakages as described in Section \ref{sec:breakMap}, we introduce a normalized metric namely \textit{breaking--to--Ref rate} as Equation \ref{eq:M3}:

\begin{equation}
\label{eq:M3}
\textit{breaking--to--Ref}~(L_v) =  \frac{|(\RC \cap \textit{\BC})|}{|\RC|} 
\end{equation}
\noindent where $L_v$ refers to a given library version. The metric $ \textit{breaking--to--Ref rate} (L_v) $ returns a percentage that ranges from [0..100\%] for each class category of $L_v$. 
Values that are closer to 100\% indicate that there are more refactorings that are breaking each of the different class categories.
Conversely, from an API breakage perspective, we now introduce a normalized metric namely \textit{Ref--to--breaking rate} to describe the ratio of overlap with respect to all breaking classes as defined in Equation \ref{eq:M2}:

%Formally, for a particular library version $L_v$, the normalized rate of refactorings that break is defined as follows:

\begin{equation}
\label{eq:M2}
\textit{Ref--to--breaking}~(L_v) = \frac{|(\textit{\RC} \cap\textit{ \BC})|}{|\textit{\BC}|} 
\end{equation}

\noindent where $L_v$ refers to a given library version. The metric $\textit{Ref--to--breaking rate}(L_v)$ returns a score that ranges from [0..1] for each class category of $L_v$. Values that are closer to 1 indicate that there are more breakages that are related to refactoring activities.

\subsection{Research Method for RQ3}
For RQ3, we used a qualitative approach to investigate the breaking APIs changes that were not detected in our approach as refactoring operations. 
%Hence, we manually investigate change logs and bug reports to cross reference against a sample of client-used API breakages that were detected as not refactorings.
Results from the prior RQ2 (See Section \ref{sec:RQ2} Table \ref{tab:RQ2a}) indicate that three of the six projects (\textsc{Guava}, \textsc{HttpClient} and \textsc{xerces}) have many client-used API breakages that were not related to refactoring activities. 
We consulted related change logs of these three projects; \textsc{Guava}\footnote{and example of Release 11 \url{https://github.com/google/guava/wiki/Release11}}, \textsc{HttpClient}\footnote{\url{https://archive.apache.org/dist/httpcomponents/httpclient/RELEASE_NOTES-4.5.x.txt}} and \textsc{Xerces} \footnote{ change logs at \url{https://xerces.apache.org/xerces2-j/releases.html}} to understand the reason why these API breakages were performed by the developer.
We manually checked documented change logs of each release to map an API breakage to either a bug fix issue or to accommodate a new enhancement (feature) in the library. 
To reduce bias, a manual check was carried by a team of three researchers (one postdoctoral and two graduate master students) persons with have an intermediate level of java programming and software development.
Since team members do not posses any project-specific knowledge, we solely rely on keywords or issues links (\ie~\texttt{issueID}) in the change log comments to map each API breakage with a bug issue or new features.
\textsc{Xerces} was later removed from the analysis as there was too many ambiguous references with no clear linkage to the source code. Analysis will include the aggregation of all API  documented changes as either bug fixes or new features and show how many can be mapped to the API breakages that did not involve any refactoring operations.

For a deeper analysis and validation, we will investigate and present some examples of these non-refactoring related API breaking classes.
%These examples will also serve as an understanding of what 

%\url{https://github.com/google/guava/wiki/} and \url{https://archive.apache.org/dist/httpcomponents/httpclient/RELEASE_NOTES-4.5.x.txt} and \url{https://xerces.apache.org/xerces2-j/releases.html}

\subsection{Research Method for RQ4}
For RQ4, we identified what refactoring operations are breaking client-used APIs.
We followed two steps in the analysis. For a library, we aggregated the number of \RC~instances where a certain \RO~(e.g., \texttt{move\_method})~has been applied. 
In the second step, we used a normalized metric $ prsv $ to describe the ratio of overlapped breaking refactorings between \activeAPI~and \inactiveAPI~as defined in Equation \ref{eq:M4}. 

\begin{equation}
\label{eq:M4}
prsv~(L, \RO)= \\  \frac{ \sum\limits_{L_v \in L}| \textit{client-used}~API\cap breaking \cap Ref~classes |}{ \sum\limits_{L_v \in L}| \textit{non client-used}~API \cap breaking \cap Ref~classes|} 
\end{equation}

\noindent where $L$ refers to a given library, \RO~ refers to a certain refactoring operation type. 

Our hypothesis is that a $\textit{prsv}$ ratio less than 1 ( 0 $\leqslant$ $\textit{prsv}$ $<$ 1) indicates that developers are applying less refactoring operations to \activeAPI. Conversely, a high $\textit{prsv}$ ratio ($ \textit{prsv} \geqslant1$ ) indicates that more refactoring operations are applied to \activeAPI. A value of 1 indicates that the certain \RO~type is equally applied to both \activeAPI~and \inactiveAPI.

%\subsection{Collected Dataset}
%\label{sec:dataset}

%\section{Results}\label{sec:extension}

%In this section, we present the results to answer our research questions. 

%For each research question, we describe (\textit{i}) the  research method, (\textit{ii}) collected data, and (\textit{iii}) the obtained results. 
 
 %\subsection{{\textit{{(RQ1). \RqOne}}}}
 
\section{Results}
\label{sec:results}
In this section, we present our results of the study by addressing each of the four research questions.

\subsection{Findings for RQ1}
\begin{figure*}[!]
\centering
	\subfloat[\textsc{Guava}]{\label{fig:Guava}%
		\includegraphics[width=.7\columnwidth]{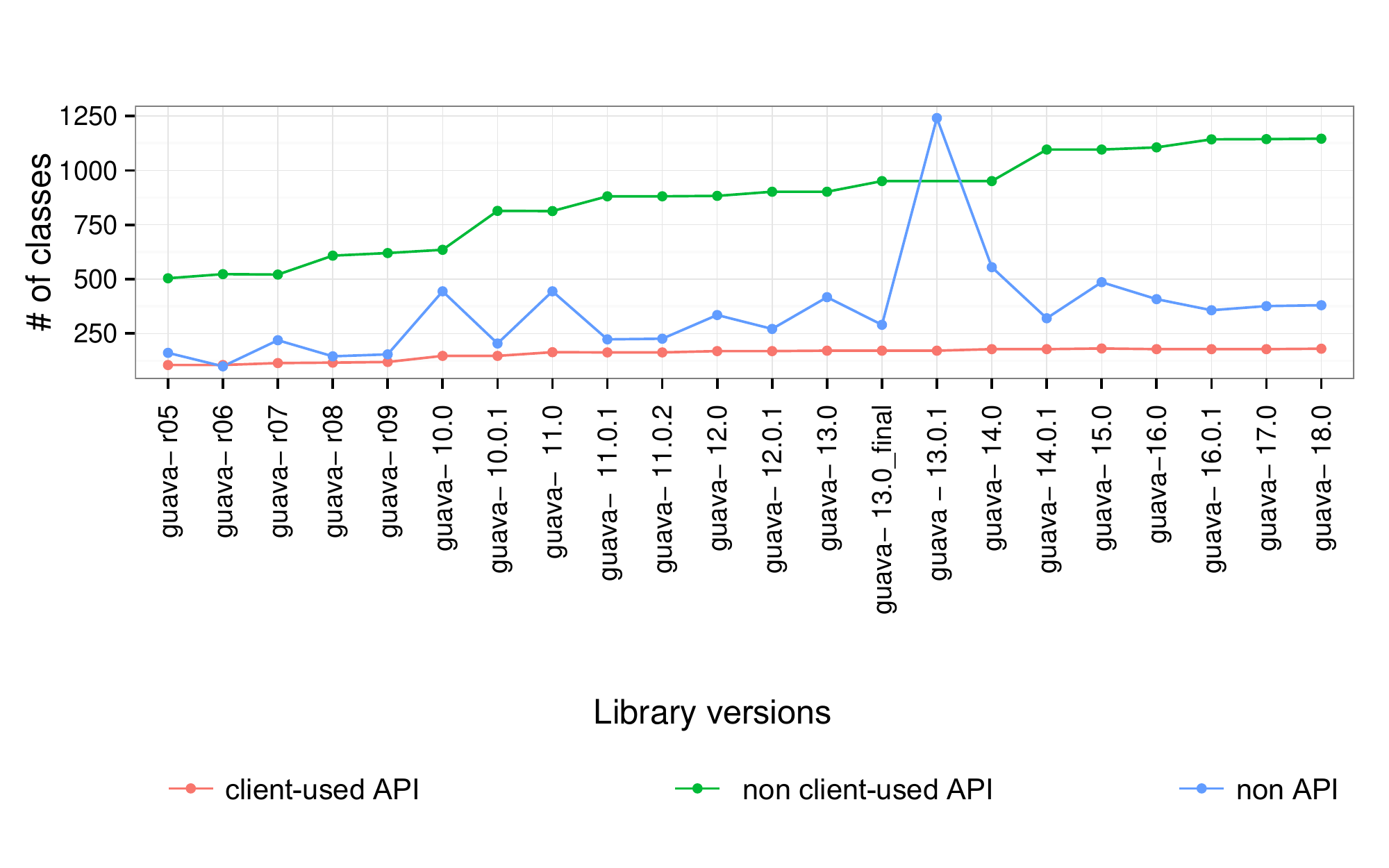}
	}\hfill
	\subfloat[\textsc{Httpclient}]{\label{fig:httpclient}
		\includegraphics[width=.7\columnwidth]{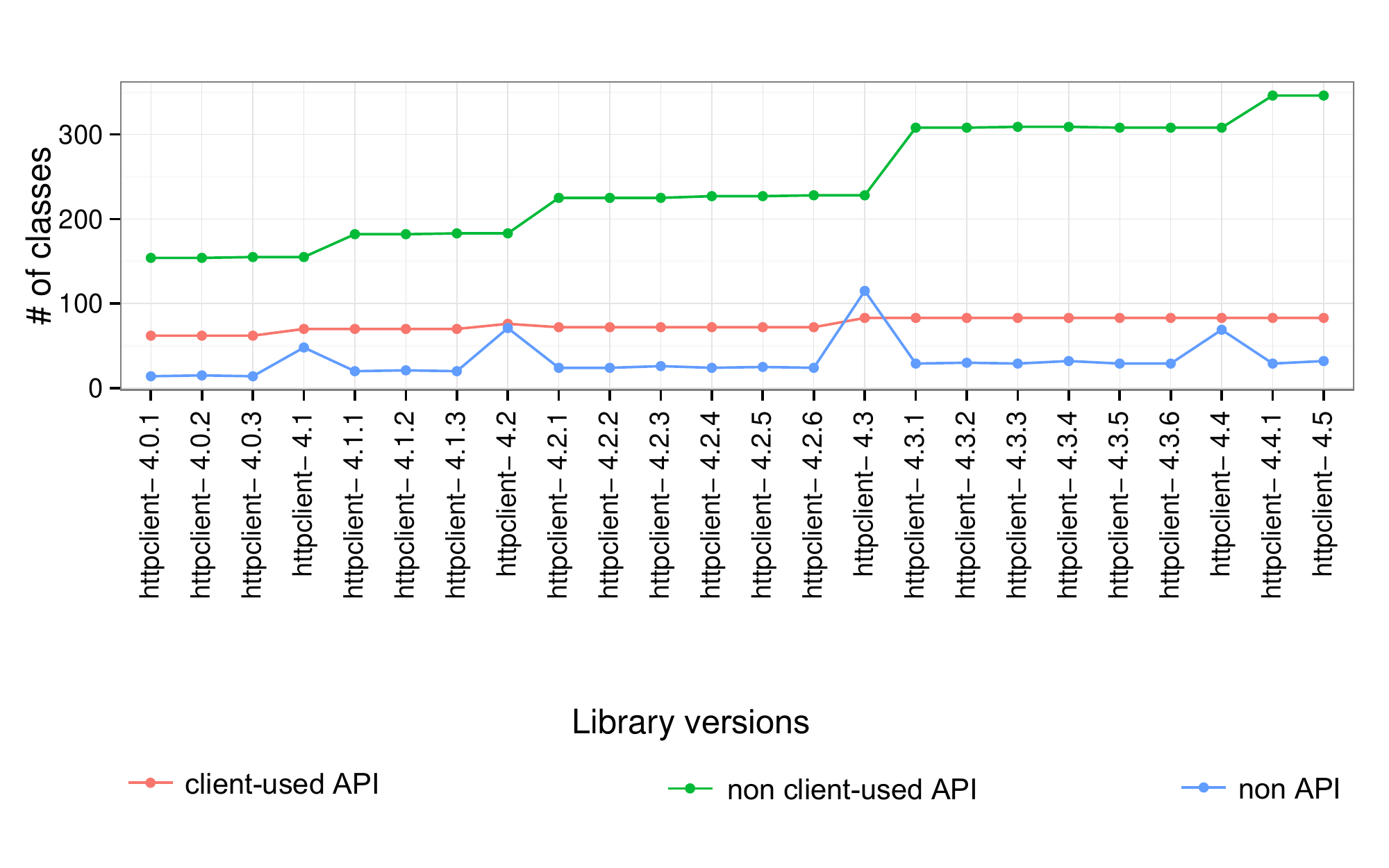}
	}\hfill
	\subfloat[\textsc{Javassist}]{\label{fig:jassist}
		\includegraphics[width=.6\columnwidth]{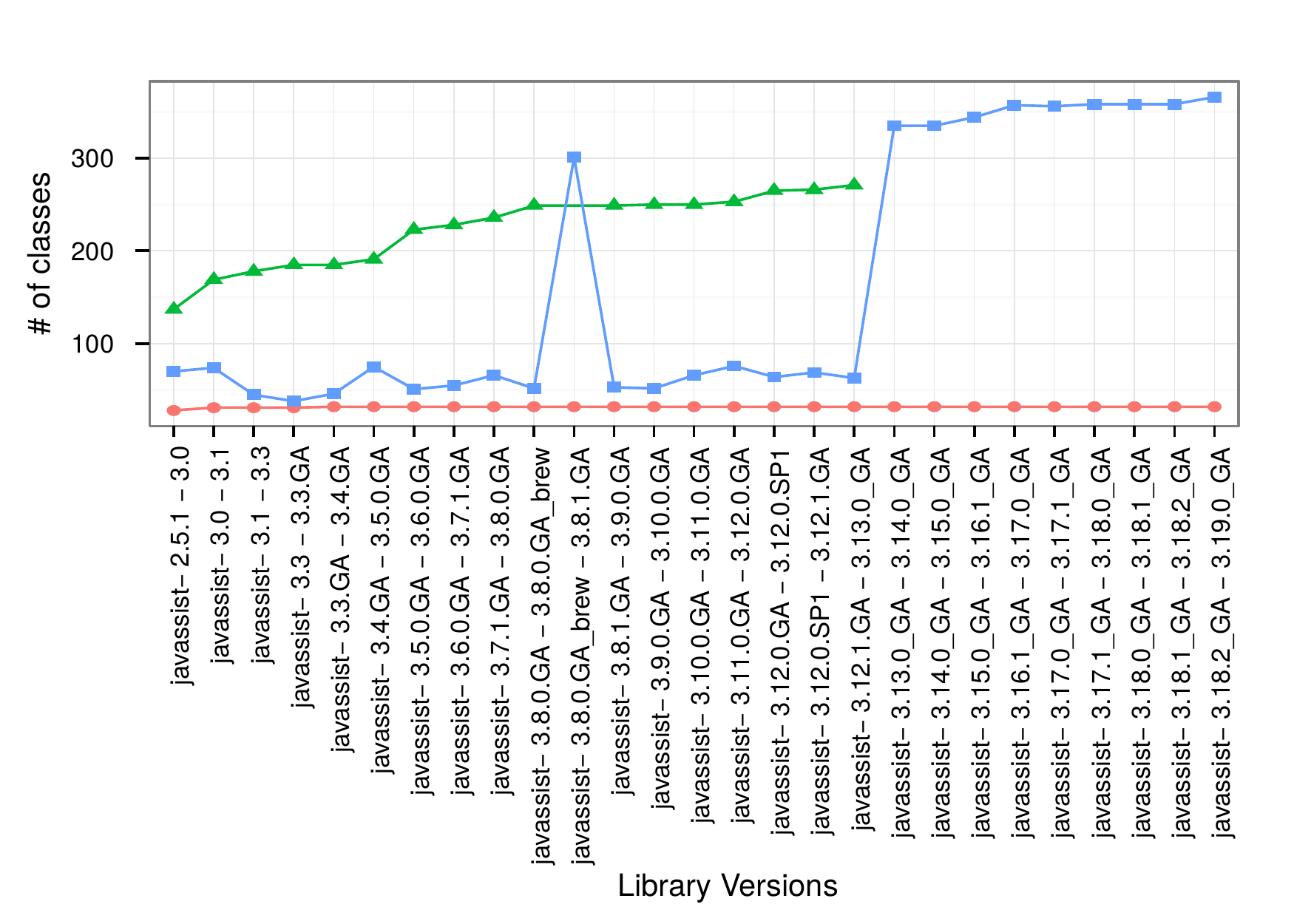}
	}\hfill
	
%\subfloat[\textsc{log4j}]{\label{fig:log4j}
%	\includegraphics[width=.8\columnwidth]{RQ1log4j}
%}
%\subfloat[$\textit{break}_{\CC}$ metric values comparing third-party API (red), internal API (green) and non API (blue) classes per  library]{\label{fig:RQ1b}
%\includegraphics[width=.8\textwidth]{RQ1b}
%}
\caption{ An evolution of changed classes per class types for (a) \textsc{guava}, (b) \textsc{httpclient}, (c) \textsc{javassist}. These figures show the different \# of classes identified in chronological order of release.}
\label{fig:RQ1}
\end{figure*}

\begin{figure*}[!]
	\centering
	\subfloat[\textsc{Jdom}]{\label{fig:jdom}
		\includegraphics[width=.7\columnwidth]{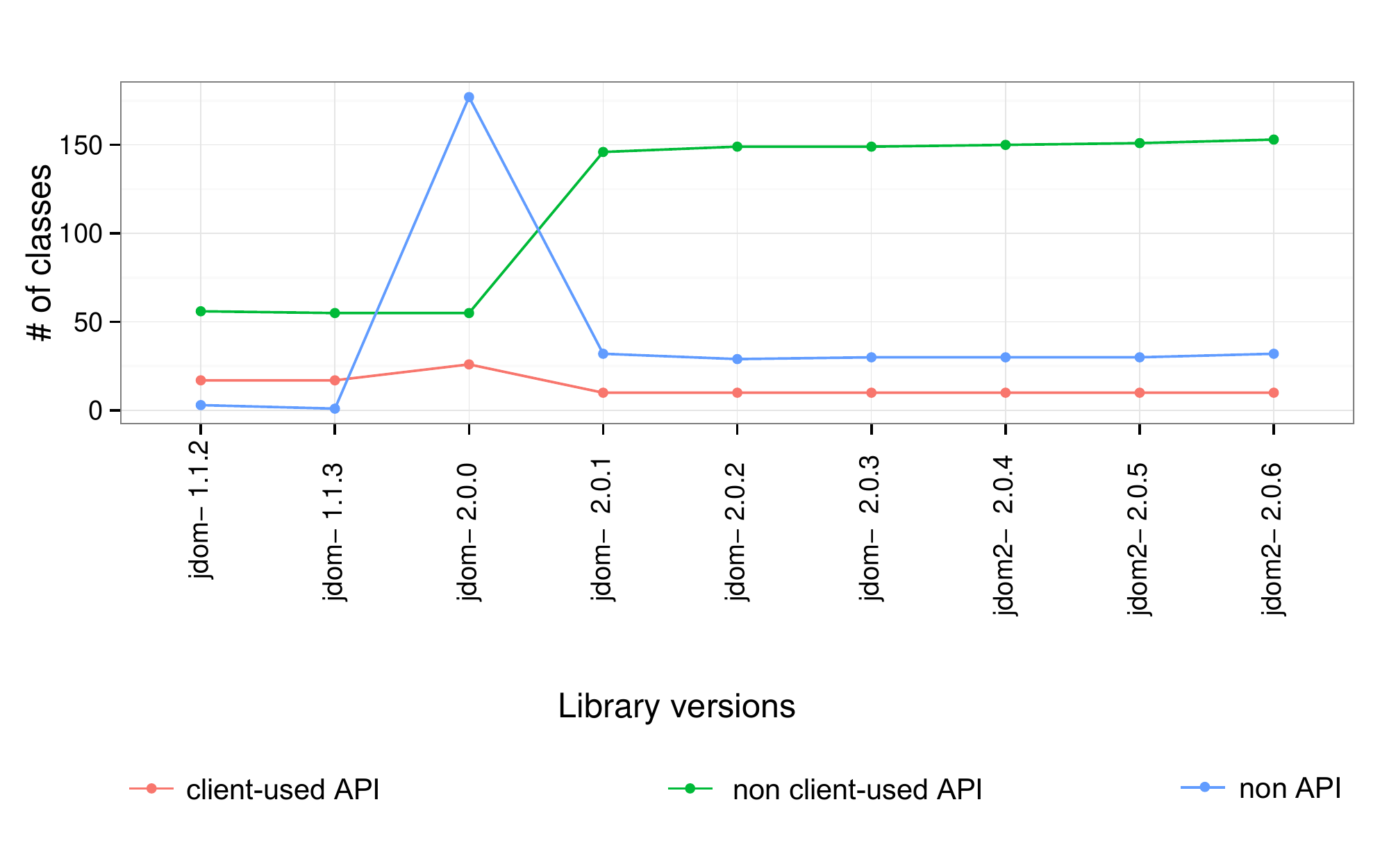}
	}\hfill
	\subfloat[\textsc{Joda-time}]{\label{fig:jtime}
		\includegraphics[width=.6\columnwidth]{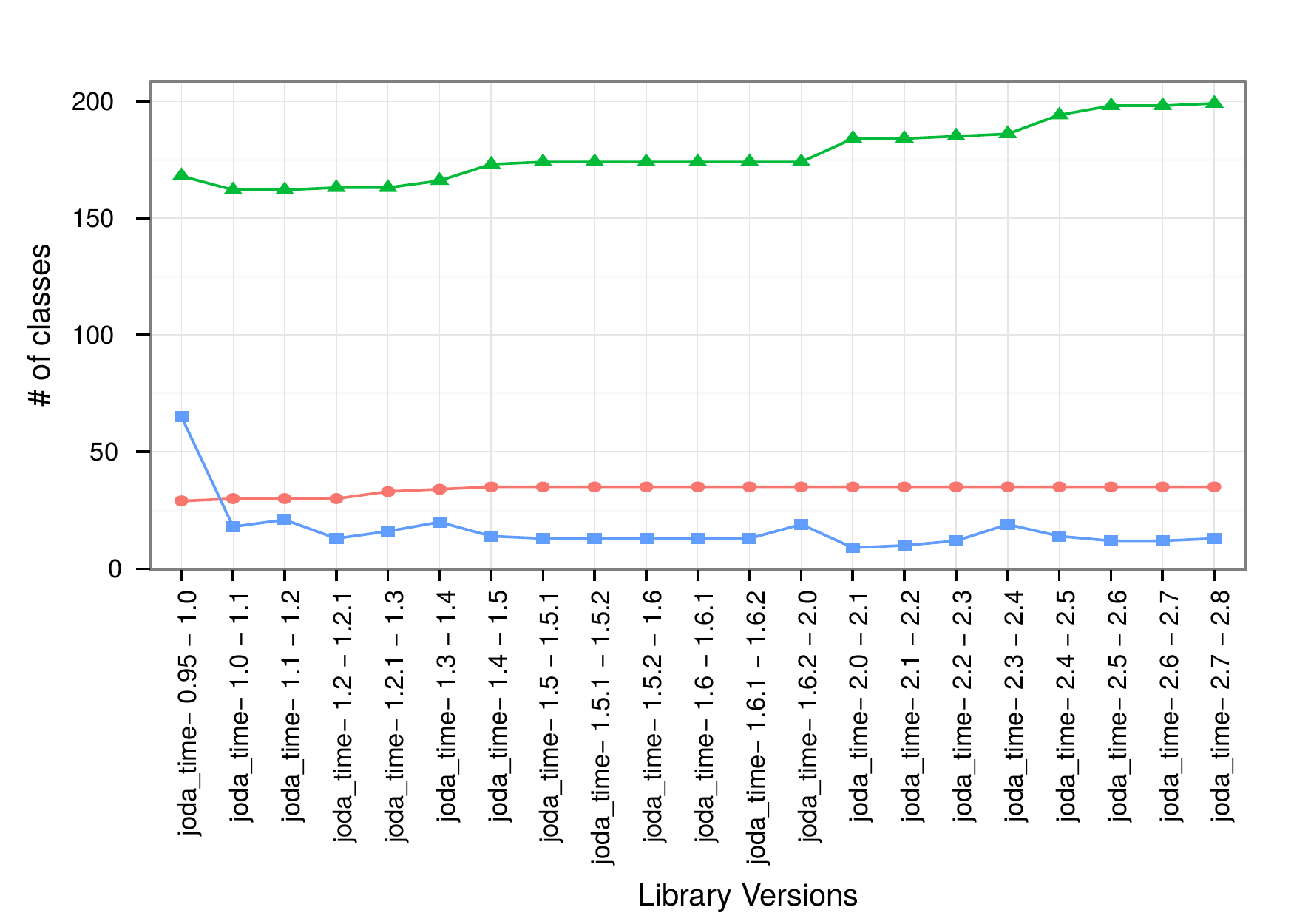}
	}\hfill
	\subfloat[\textsc{log4j}]{\label{fig:log4j}
		\includegraphics[width=.7\columnwidth]{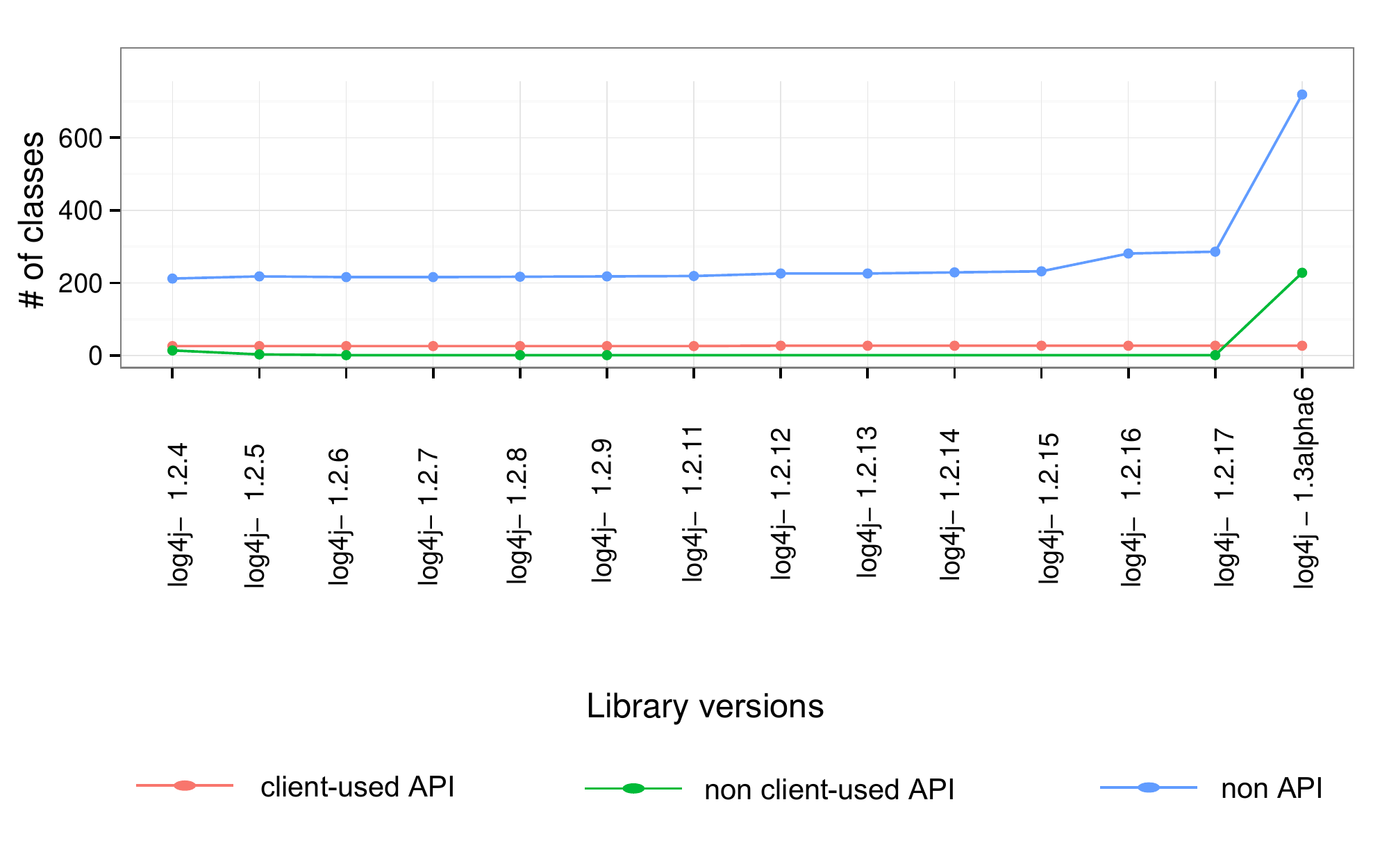}
	}\hfill

	%\subfloat[$\textit{break}_{\CC}$ metric values comparing third-party API (red), internal API (green) and non API (blue) classes per  library]{\label{fig:RQ1b}
	%\includegraphics[width=.8\textwidth]{RQ1b}
	%}
	\caption{ An evolution of \# of classes per class types for (a) \textsc{jdom}, (b) \textsc{joda-time} and (c) \textsc{log4j} libraries. Similar to Figure \ref{fig:RQ1}, these figures show the different \# of classes identified in chronological order of release.}
	\label{fig:RQ11}
\end{figure*}

\begin{figure*}[!]
	\centering
	\subfloat[\textsc{Slf4j}]{\label{fig:slf4j}
		\includegraphics[width=.9\columnwidth]{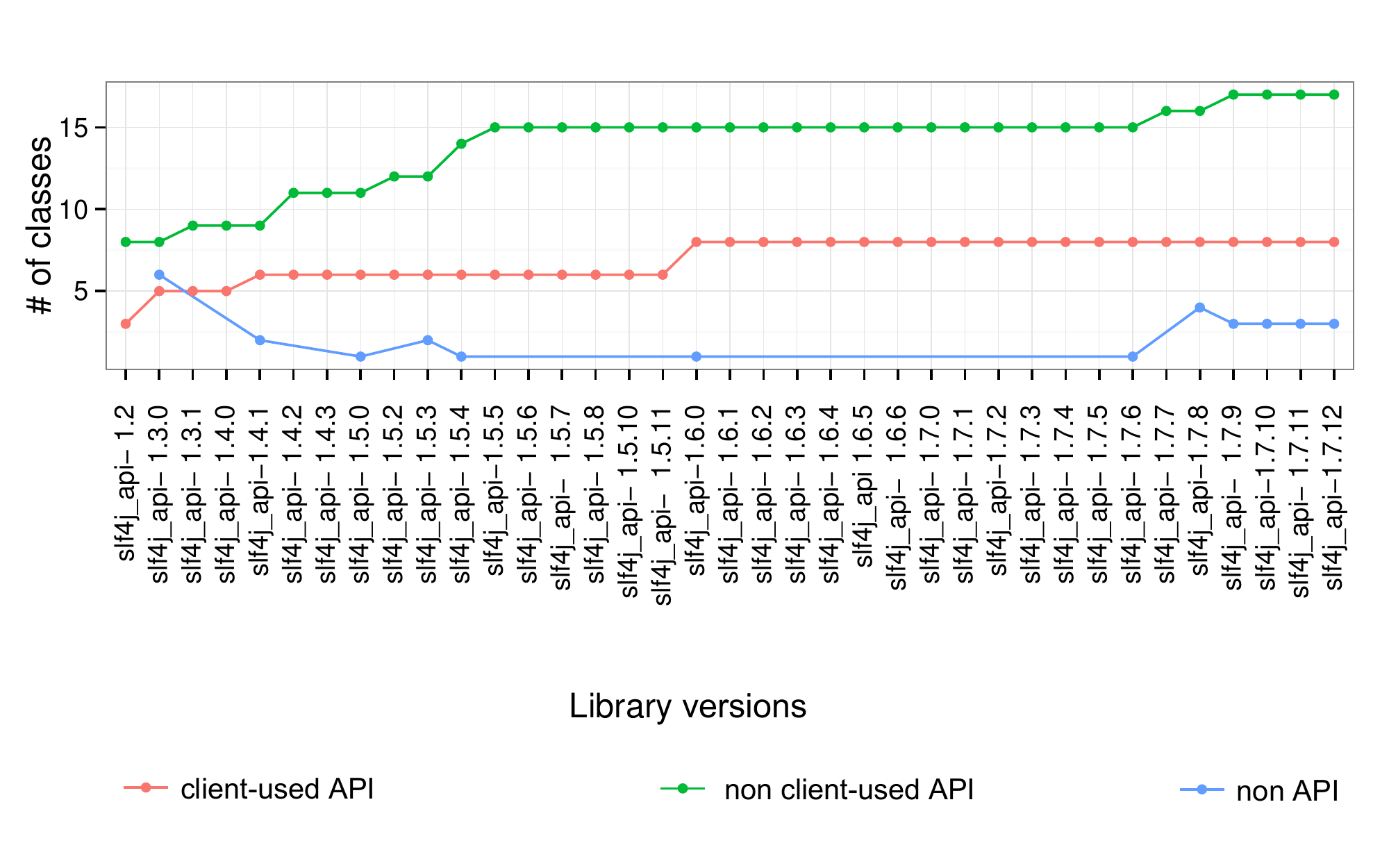}
	}\hfill
	\subfloat[\textsc{Xerces}]{\label{fig:xerces}
		\includegraphics[width=.9\columnwidth]{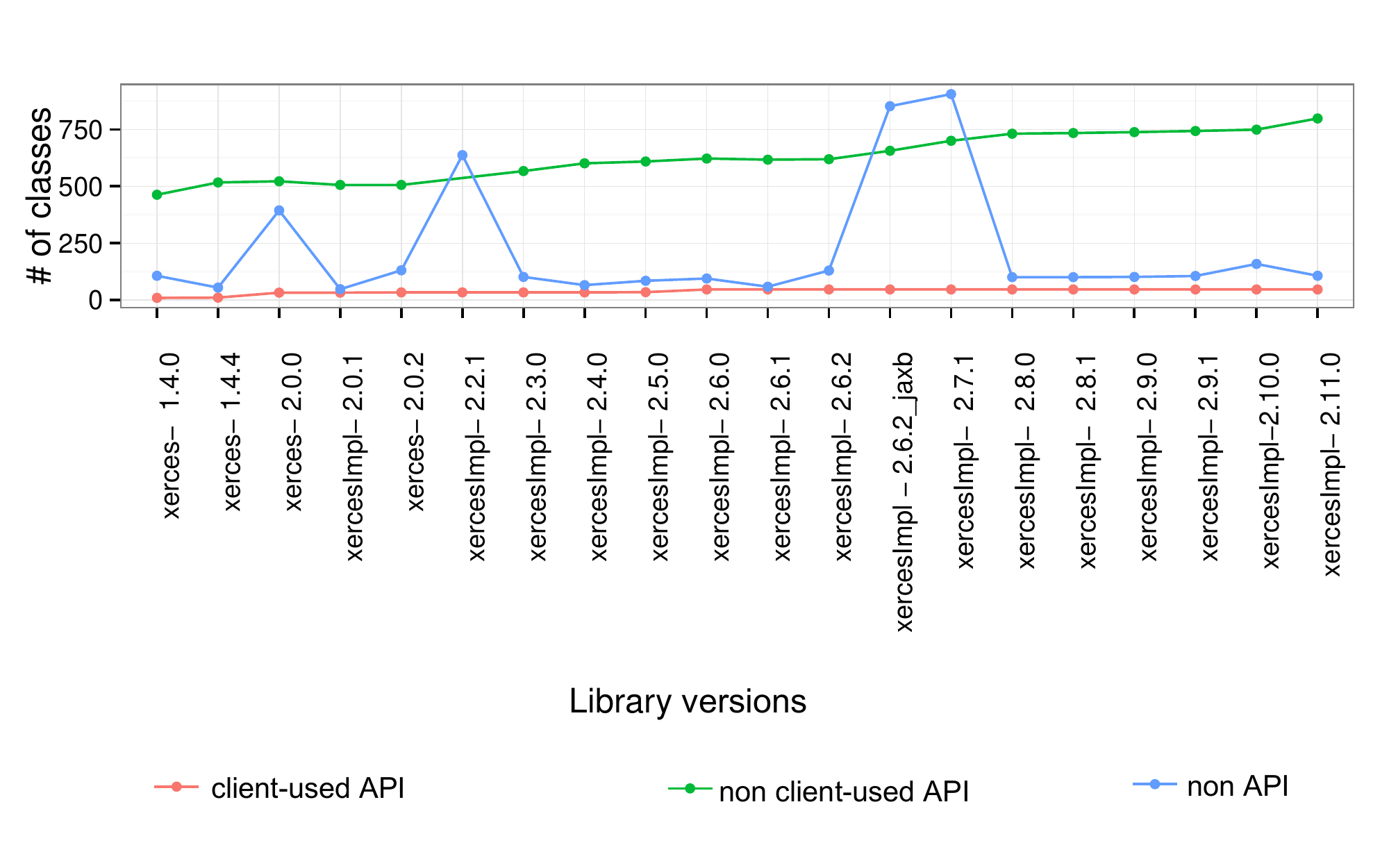}
	}\hfill
	\caption{ An evolution of \# of classes per class types for (a) \textsc{slf4j} and (b) \textsc{xerces} libraries. Similar to Figure \ref{fig:RQ1} and Figure \ref{fig:RQ11}, these figures show the different \# of classes identified in chronological order of release.}
	\label{fig:RQ12}
\end{figure*}

Figures \ref{fig:RQ1}, \ref{fig:RQ11} and \ref{fig:RQ12} depict class category analysis of each consecutive library version. 
Each figure shows the evolution of (i) \activeAPI, (ii)  \inactiveAPI~and (iii) \nonapi~over consecutive library versions. 
From these figures, we summarize our findings with three observations (\ie~Figures \ref{fig:Guava}$\sim$\ref{fig:xerces}). 
First, we observe that most libraries are composed of \inactiveAPI~categories (green line), showing that libraries usually have more \inactiveAPI~than \activeAPI.
The exception is \textsc{log4j}, which is shown in Figure \ref{fig:log4j} to have most APIs intended for external API usage. Interestingly, we see in Figure \ref{fig:jassist} that  \inactiveAPI~of  \textsc{javassist} disappears from the more recent libraries. 
Upon closer inspection, we noticed that this was because developers had changed these \inactiveAPI~into \nonapi. 
Second, we observe a stable number of \activeAPI~(red line) shown across all projects.
From a client user viewpoint, the findings indicate that developers of a library are less likely to expand their external APIs. 
The obtained results show that the number of \nonapi~(blue line) is constantly changing (\ie~illustrated by various peaks) over time. 
We find that some of the peak changes can be correlated to different events, such as a major or specially-named releases,  beta releases such as \texttt{xercesImpl2$_{.x.x\_jaxb}$} and \texttt{log4j$_{1.3alpha6}$}, or modifying private \nonapi~into public APIs such as in the case of  \textsc{httpclient}. 

\begin{figure*}
	\centering
		\subfloat[{For each of the eight libraries, we show the $\textit{break}_{\CC}$ comparing (1) \activeAPI~in red, (2) \inactiveAPI~in green and (3) \nonapi~in blue.}]{\label{fig:RQ1b}
			\includegraphics[width=.8\textwidth]{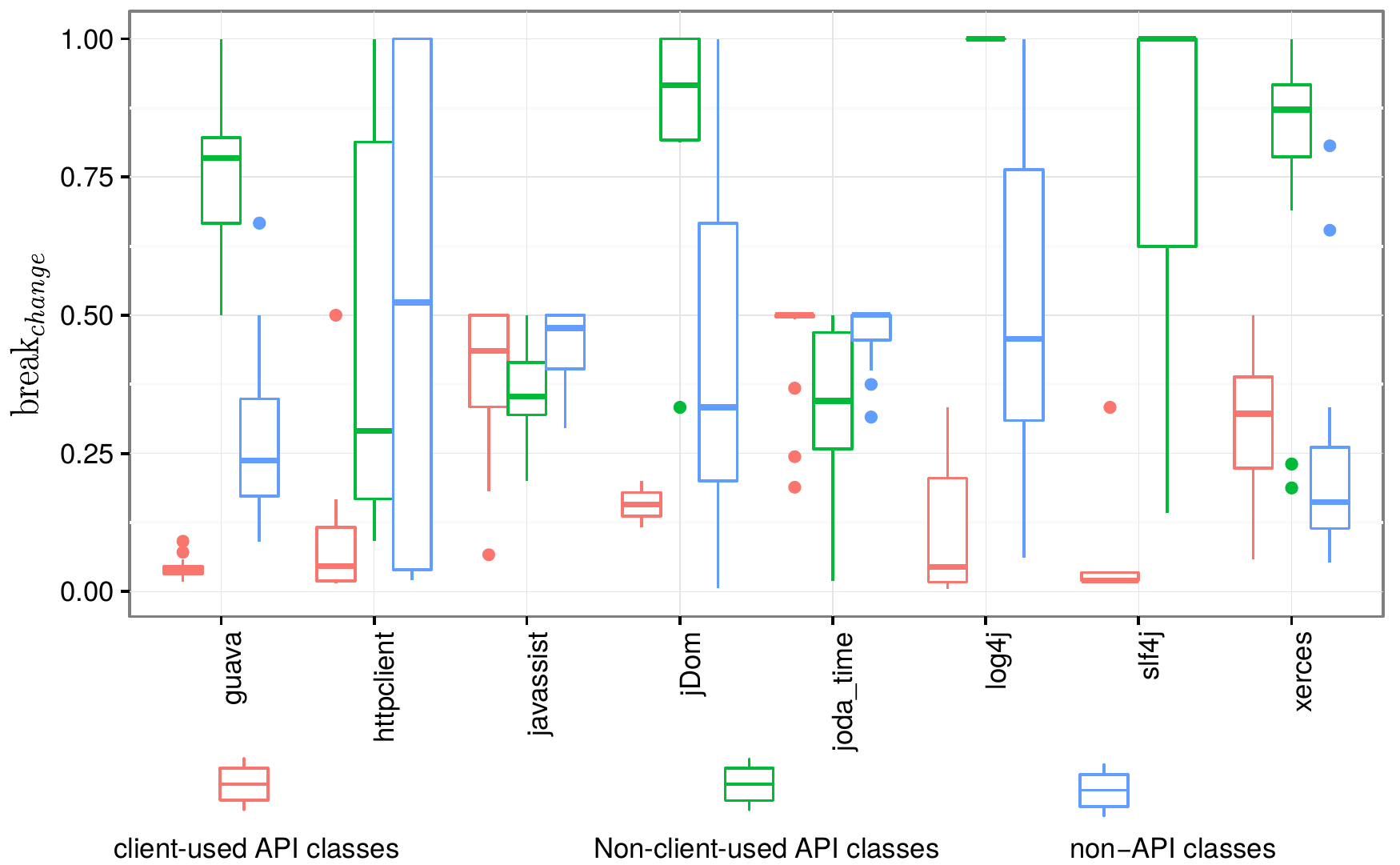}
		}\hfill
	\subfloat[ Corresponding to Figure \ref{fig:RQ1b}, this table shows the \# of analyzed (1) library versions, (2) breaking and (3) changed classes collected. ]{\small
	\begin{tabular}{@{}lcrrr@{}}
		\toprule
		& \# versions &breaking class instances& changed class instances \\ \midrule
		\textsc{Guava}  & 22& 2,215 & 9,973\\
		\textsc{httpclient}& 25&113& 1,426\\
		\textsc{Javassist}& 28 &1,017& 2,572 \\
		\textsc{Jdom}& 10& 106& 445\\
		\textsc{Joda-time}&22 &1,097 & 2,922\\
		\textsc{log4j}& 17& 583& 3,051\\ 
		\textsc{Slf4j}  & 38& 21& 235 \\
		\textsc{Xerces}& 21& 4,622& 7,796 \\
		\bottomrule
		Totals &183& 9,774& 28,420\\
	\end{tabular}}
		\caption{Results of the $break_{change}$ rates for all eight libraries analyzed.\label{f:main}}
\end{figure*}

\begin{figure}
	\centering
	\includegraphics[width=.6\textwidth]{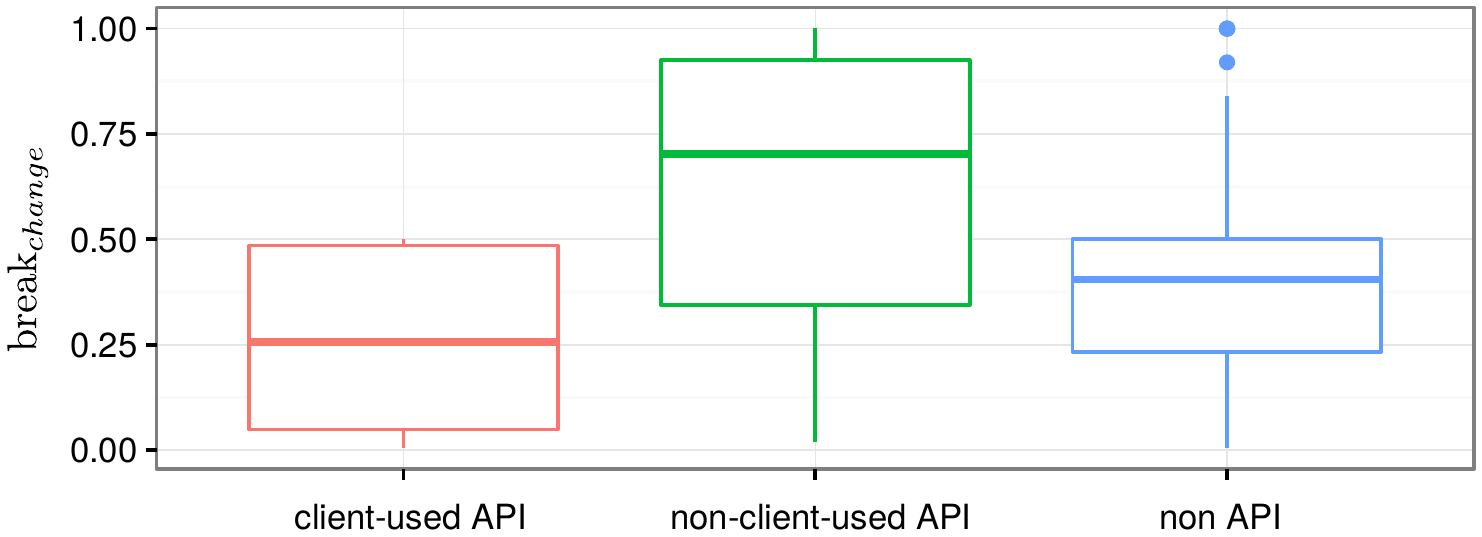}
	\caption{Summary of \textit{break}$_{\CC}$ comparing (1) \activeAPI~in red, (2) \inactiveAPI~in green and (3) \nonapi~in blue.}
	\label{fig:RQ1bb} 
\end{figure}

\begin{hassanbox}
\label{sec:RQ1aSummary}
Library maintainers are less likely to apply \activeAPI~changes compared to other class categories. 
%We observe that code additions and removals to \inactiveAPI~and \nonapi~are more frequent than \activeAPI.
%Also, the number of \activeAPI~is consistent over time.
%The results suggesting that clients do not seem to expand beyond the set of \activeAPI. 
\end{hassanbox}

%\begin{table}
%	\fontsize{7}{7}\selectfont
%	%	\centering
%	\caption{Library Class Categories Incompatibility Matrix}.
%	\label{tab:breakTypes}
%	\begin{tabular}{l|c|cc}
%		\toprule
%		\multicolumn{1}{c}{} & \multicolumn{1}{c}{Compatible Change} & \multicolumn{2}{c}{Incompatible Changes ($\textit{break}_{change}(L_v)$)}\\
%		\midrule
%		\cellcolor{red!25} \textsc{client-used API}&API compatible change &\cellcolor{red!25} API Breaking code change\\
%		\cellcolor{green!25} \textsc{non client-used API} & API compatible change &Incompatible change unintended for client \\
%		\cellcolor{cyan!50} \textsc{non API}& Change does not affect client& Incompatible change does not affect client\\ 
%		\bottomrule
%	\end{tabular}
%\end{table}

Figure \ref{fig:RQ1b} shows the ${break}_{change}$ rates for all eight libraries. From this figure, we observe that except for \textsc{javassist} and \textsc{joda-time}, library developers are more likely to break \inactiveAPI~than \activeAPI.
%In the case of\textsc{slf4j}, the reason might be due to the absence of breakages found in its \nonapi.
Related, Figure \ref{fig:RQ1bb} depicts the ${break}_{change}$ rates grouped by all class categories. The Figure shows that \inactiveAPI~are more prone to breakages than  \activeAPI~for all libraries. 
As shown in the Figure, \inactiveAPI~are reported to have the most breakages.
A Kruskal Wallis test revealed a significant differences between \activeAPI, \inactiveAPI~and \nonapi~values (p$<$0.01). 
The post-hoc test using Mann-Whitney tests with Bonferroni correction proves the effect size to be medium  (p$<$0.01, r = 0.54) when comparing all class categories.

\begin{hassanbox}
\label{sec:RQ1bSummary}
Findings show that incompatible API code changes are statistically more likely to occur in \inactiveAPI~compared to \activeAPI. 
%The obtained results provide statistical evidence that library maintainers are more likely to break other classes than client-used APIs.
\end{hassanbox}

\begin{table*}[t] \centering 
	\fontsize{10}{10}\selectfont
	\tabcolsep=0.1cm
	\caption{The table reports (a) number of \texttt{Ref classes} and (b) \RO~density per Ref class ($\bar{x}=|$\RO$|$). Note that (--) represents no matches.} 
	\label{tab:RQ2} 
	\begin{tabular}{@{\extracolsep{2pt}} llccccccc} 
		\toprule 
		&&  \multicolumn{3}{c}{$breaking~classes$} & \multicolumn{3}{c}{$non~breaking$}\\ 
		\cmidrule(r){3-5} \cmidrule(r){6-8}\\
		
		& &\rotatebox[origin=c]{70}{\textit{\activeAPIAB}}  &\rotatebox[origin=c]{70}{\textit{\inactiveAPIAB}}   &\rotatebox[origin=c]{70}{\textit{non API}} &  \rotatebox[origin=c]{70}{\textit{\activeAPIAB}}  &\rotatebox[origin=c]{70}{\textit{\inactiveAPIAB}}   &\rotatebox[origin=c]{70}{\textit{non API}} \\ 
		%  & ($\bar{x}=\RO$ per class) & ($\bar{x}=\RO$ per class) &($\bar{x}=\RO$ per class) & ($\bar{x}=\RO$ per class) & ($\bar{x}=\RO$ per class) & ($\bar{x}=\RO$ per class)  \\
		 \midrule\\
%		  & & &  &  &  & & \\
		$|Ref|$  &{\textsc{guava}} & {32} & 143 & 44 & 31 & 24 & 139 \\ 
		 &\textsc{httpclient} & 3 & 11 &-- & 6 & 16 & 7 \\ 
		&\cellcolor{gray!25}\textsc{Javassist}& \cellcolor{gray!25}{111} & 8& 6 & 29 & 10 & 2  \\ 
		&\textsc{jdom} & 1 & 1 &-- &-- & 3 &-- \\ 
		&\textsc{Joda-time}& 30  & --  &11 & 29 & 5 & 12\\ 
		&\textsc{log4j} & 1 &-- &-- & 2 &--  & 3 \\ 
		&\textsc{slf4j} &--&1 &--  & 2 & 1 &-- \\ 
		&\cellcolor{gray!25}{\textsc{xerces}} & 44 & \cellcolor{gray!25} {244} & 31 & 23 & 104 & 66 \\
		\hline \\ [-1.8ex] 
		$\RO$ density  & & &  \\
		(Median) &\cellcolor{gray!25}{\textsc{guava}} & \cellcolor{gray!25} {7} & 8 & 5 & 4 & 4 & 8 \\ 
		&\textsc{httpclient} & 5 & 5 &-- & 2 & 3 & 2 \\ 
		&\textsc{Javassist}& 2 & 5 & 2 & 1& 1& 1\\ 
		&\textsc{jdom} &1 & 1 &-- &-- & 4 &-- \\
		&\textsc{Joda-time}& 3 & -- & 10&1 &1 &1\\  
		&\textsc{log4j} & 1 &-- &-- & 2 &--  & 2 \\ 
		&\cellcolor{gray!25}{\textsc{slf4j}}&--&1 &--  & 3 & \cellcolor{gray!25} {22} &-- \\ 
		&\textsc{xerces} &8 & 6 & 4 & 4 & 4 &5 \\
		\bottomrule 
	\end{tabular} 
\end{table*} 

%\begin{landscape}
	
\begin{table*}[t] 
	\fontsize{10}{10}\selectfont
	\centering 
	\tabcolsep=0.1cm
	\caption{Matrix that shows the median average \# refactored API classes per library. For each library, we summarized the median values across all library versions. Table includes median ($\bar{x}$) of matched refactored classes. 0 represents a value less than 0.01. (--) reports no matched classes. } 
	\label{tab:RQ2a} 
	\begin{tabular}{@{\extracolsep{5pt}} lcccccccc} 
		\toprule
		& & & \multicolumn{2}{c}{$|Ref|$}  &  \multicolumn{2}{c}{$|non~Ref|$}
		&\multirow{5}{*}{\rotatebox[origin=c]{70}{\textit{Ref--to--breaking rate}}} &  \multirow{5}{*}{\rotatebox[origin=c]{70}{\textit{breaking--to--Ref rate}}}   \\ 
		& & & \multicolumn{2}{c}{(Median)}  &  \multicolumn{2}{c}{(Median)}\\
		\cmidrule(r){4-5} \cmidrule(r){6-7}\\
		&&& \multirow{3}{*}{\rotatebox[origin=c]{70}{breaking} } &  \multirow{3}{*}{\rotatebox[origin=c]{70}{non breaking} }  &  \multirow{3}{*}{\rotatebox[origin=c]{70}{breaking} } &  \multirow{3}{*}{\rotatebox[origin=c]{70}{non breaking} }\\
		& &\rotatebox[origin=c]{70}{\textit{\# versions}}\\ [-1.8ex]\\ 
		\midrule
		\multirow{8}{*}{\rotatebox[origin=c]{90}{\activeAPIAB}}   &\cellcolor{gray!25}{\textsc{guava}}& 22 & \cellcolor{gray!50} $\textbf{2}$ & $2$ & $166$ & $251$ & $1\%$ & $53\%$\\ 
		&\textsc{httpclient} & 25 & $1$ & $1$ & $46$ & $74$& $1\%$ & $55\%$ \\ 
		&\textsc{Javassist}& 28  & 2 & 1 &3 & 1 & \cellcolor{gray!25} \textbf{37\%} & $\cellcolor{gray!25} \textbf{75\%}$ \\ 
		&\textsc{jdom} & 10& $1$ & $0$ & $1$ & $16$ & $1\%$ & $1\%$   \\
		&\textsc{Joda-time}& 22 & 15 & 2 & 65 & 45 & 6\% & 48\% \\  
		&\textsc{log4j}& 17 & $0$ & $1$ & $87$ & $75$ & $0\%$ & $0\%$ \\ 
		&\textsc{slf4j}&38 & $-$ & $1$ & $41$ & $136$ & $0\%$ & $0\%$ \\ 
		&\textsc{xerces}&21 &  $3$ & $2$ & $24$ & $16$ &$10\%$ & 64\% \\
		\midrule\\		 
		 \multirow{8}{*}{\rotatebox[origin=c]{90}{\inactiveAPIAB}}  
		 &\cellcolor{gray!25}{\textsc{guava}} & 22& \cellcolor{gray!25} {$9$} & $2$ & $91$ & $20$ & $9\%$ & \cellcolor{gray!25} \textbf{86\%}\\ 
		&\textsc{httpclient} & 25& $2$ & $1$ & $8$ & $18$ &\cellcolor{gray!25}\textbf{14\%} & $58\%$ \\ 
		&\textsc{Javassist}& 28 & 4 & 1 & 2 & 1  &5\% &44\%\\  
		&\textsc{jdom} & 10& $1$ & $1$ & $13$ & $3$ & $1\%$ & $22\%$  \\ 
		&\textsc{Joda-time}& 22 & -- & 1 & 9 & 7 &-- &--\\ 
		&\textsc{log4j} & 17& $0$ & $-$ & $36$ & $-$& $0\%$ & $-$\\ 
		&\textsc{slf4j}  & 38& $0$ & $1$ & $2$ & $2$& $0\%$ & $0\%$  \\ 
		&\textsc{xerces} & 21& $14$ & $7$ & $210$ & $22$& $6\%$ & $68\%$ \\ \bottomrule
	\end{tabular} 
\end{table*}

\subsection{Findings for RQ2}
\label{sec:RQ2}

Table \ref{tab:RQ2} presents a summary of \texttt{Ref classes} and their \RO~density. 
For instance, we identified 32 \textsc{guava} \activeAPI~that were \texttt{Ref classes}.
Out of the 32 \texttt{Ref classes}, we report a median of 7 \RO~that were applied per \texttt{Ref class}. 
From this table, we can see that, in general, library maintainers applied more \RO~to \inactiveAPI~and \nonapi, as compared to \activeAPI, except for \textsc{javassist}.
For example, the table shows that for \textsc{xerces}, around 244 \inactiveAPI~were refactored, compared to 44 \activeAPI. 
In more detail, the results show that apart from \textsc{slf4j}, the median density of \RO~per class ranges from 1 to 10 \RO~at most. Interestingly, we find that \textsc{slf4j} had a high number of \RO~applied to one non-breaking \activeAPI~($\bar{x}$=22).
\textsc{log4j}, \textsc{slf4j} and \textsc{jdom} libraries reported only a few breaking classes matched to \RO, which is consistent with recent empirical studies conducted by Cossette \textit{et al.} \cite{Cossette2012}. 
 
Table \ref{tab:RQ2a} reports the median values of \RO~that cause API breakages. 
We use this table to compare between \activeAPI~and \inactiveAPI.
For \textsc{guava}, \inactiveAPI~($\bar{x}$=9) were breaking due to refactorings compared to \activeAPI~($\bar{x}$=2).
From this table, we find that non refactoring changes are more likely to break \activeAPI~than \inactiveAPI. 
Moreover, applied refactorings tend to break more \inactiveAPI~compared to \activeAPI.   
The results show that many of the API breakages are not mapped to the detected refactorings (\ie~\texttt{non Ref classes}).
% reported more breakages to \activeAPI~compared to \inactiveAPI. 
%For example, the Table shows that non Ref classes that are more likely to break than  ($\bar{x}$=165.57, 45.57, 70.50, 187.14 and 41). 
We find that more refactoring \inactiveAPI~are breaking compared to refactored \activeAPI, with the exception of \textsc{Javassist}. 

Table \ref{tab:RQ2a} also shows the \textit{breaking--to--Ref} and \textit{Ref--to--breaking} rates. % for the refactorings that were breaking API classes. 
We report that the median \textit{Ref--to--breaking rate} for \activeAPI~is up to 37\% across all projects ($\bar{x}$=1\%$\sim$37\%).
Except for \textsc{javassist}, the result provides evidence the detected API breakages could not be mapped to refactoring operations.
%It is, however, important to note that \textit{javassist}
%In fact, we observe the highest median of 14\%  were instead related to \inactiveAPI.  
Alternatively, the \textit{breaking--to--Ref rates} reported for \activeAPI~in Table \ref{tab:RQ2a} indicates that breaking refactorings accounted for a median range of up to 75\% of all \RO. 
%Comparatively, prior work \cite{Dig2006} reported much higher values (up to 97\%).
The highest \textit{breaking-Ref} rate for \inactiveAPI~was 86\%, reported for the \textsc{guava} library. 

%However, it is worth to notice that our results match well with Cossette \textit{et al.} \cite{Cossette2012}, it do not much well with Dig and Johnson \cite{Dig2006} who found that API breakages are more related to refactoring activities. We address this in Section \ref{sec:related}.

%\textcolor{red}{In fact, we assume that the inclusion of undocumented API breakages may be a cause for differences to prior work. Due to the size of our dataset, we consider that a more in-depth investigation of this phenomena is regarded as separate future work.}

\begin{hassanbox}
\label{sec:RQ2SumA}
Findings show that up to 75\% refactored API classes are breaking their client-used APIs. However, these API breaking refactorings account for less than 37\% of all client-used API breakages.
\end{hassanbox}

\subsection{Findings for RQ3}

\begin{table}
	\fontsize{10}{10}\selectfont
	\centering 
	\tabcolsep=0.1cm
	\caption{Shows for a library release (i) the number of issues and new features per version release analysis and (ii) the number of non-refactoring related API breaking classes. We also show in the number of these API breakages mapped to the change log comments. } 
	\label{tab:RQ33} 
	\begin{tabular}{@{\extracolsep{3pt}} lrccccc } 
		\toprule	
		Library&Change Log &\multicolumn{1}{c}{\# Issues} &  \multicolumn{1}{c}{\# New Features} & $|non~Ref| \cap$ breaking\\
		&Release&&&$\cap$ \activeAPIAB \\
		&&  &&(\# mapped to change logs)\\   		
		\hline \\[-1.8ex] 
		\cellcolor{gray!25} \textsc{Guava}&\cellcolor{gray!25} v11 &\cellcolor{gray!25}  26 & \cellcolor{gray!25} 13 &\cellcolor{gray!25}4 (4) \\ 
		&v12& 43 & 24 &4 (4)  \\ 
		&v13& 26 & 28 &7 (6)  \\ 
		&v14& 64 & 10 &5 (5)\\ 
		&v15 & 53 & 11 &7 (6)  \\ 
		&v16 & 19 & 8 &5 (4)\\ 
		&v17 & 11 & 5 &6 (4)\\ 
		&v18 & 21 & 7 &3 (2)\\ \hline \hline
		& &263&106&41 (34) 82\%\\ \hline
		\textsc{Httpclient} &v4.1.2 &5&-&3 (2)\\
		&v4.1.3 &4&-&3 (2)\\
		&v4.2  &15&4&3 (1)\\
		&v4.2.1 &8&-&4 (2)\\
		&v4.2.2 &8&-&4 (2)\\
		&v4.2.3 &21&-&2 (2)\\ 
		&v4.2.4 &6&9&3 (1)\\
		&v4.2.5 &6&4&2 (2)\\\hline \hline
		& &73&17&24 (14) 58\%\\ \bottomrule
		\\[-1.8ex] 
	\end{tabular} 
\end{table} 

Table \ref{tab:RQ33} shows results of the manual study  of API breakages that did not map to any \texttt{Ref Classes} in developer documentation (\ie~change logs). 
For instance, release 11 of the \textsc{guava} library listed 26 issues and 13 new features\footnote{The release notes are available at \url{https://github.com/google/guava/wiki/Release11}}.
The table confirms our results that find these client-used API class breakages are not only related to refactoring activities (\ie~ \texttt{non Ref Class}). 
We find that all four API breakages could be mapped to the API documented changes. 
From the table, we were able to map 82\% of non refactoring client-used API breakages to the API documentation for \textsc{guava} and 58\% for \textsc{httpclient}. 
This finding indicates that many of the API breakages not involved in refactorings are most likely motivated by maintenance issues such as bug fixes and for new feature enhancements.
%We find evidence of many issues and new feature fixes during the evolution of the studied libraries.

\begin{figure}
		\centering
	
\subfloat[This code change in the method \textsc{resolve} was detected as breaking API compatibility for users of the older JDK.  ]{\label{fig:RQ3ex1} 
		\includegraphics[width=0.9\textwidth]{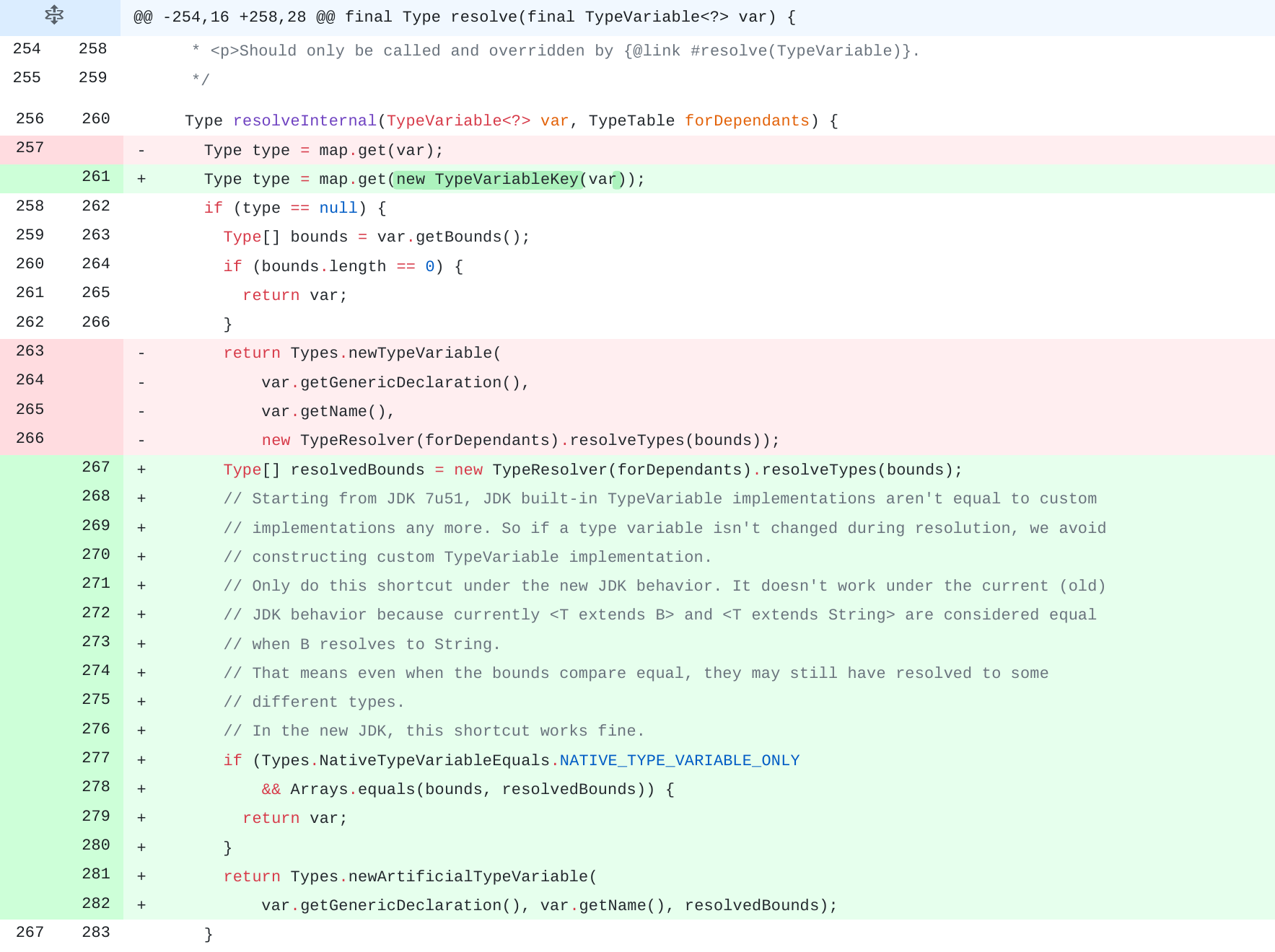}
	}\hfill
	\subfloat[This code change breaks API compatibily with the replacement \textsc{HashCodes}  to \textsc{HashCode} in the method. This refactoring was missed by the automated approach]{\label{fig:RQ3ex2} 
		\includegraphics[width=0.9\textwidth]{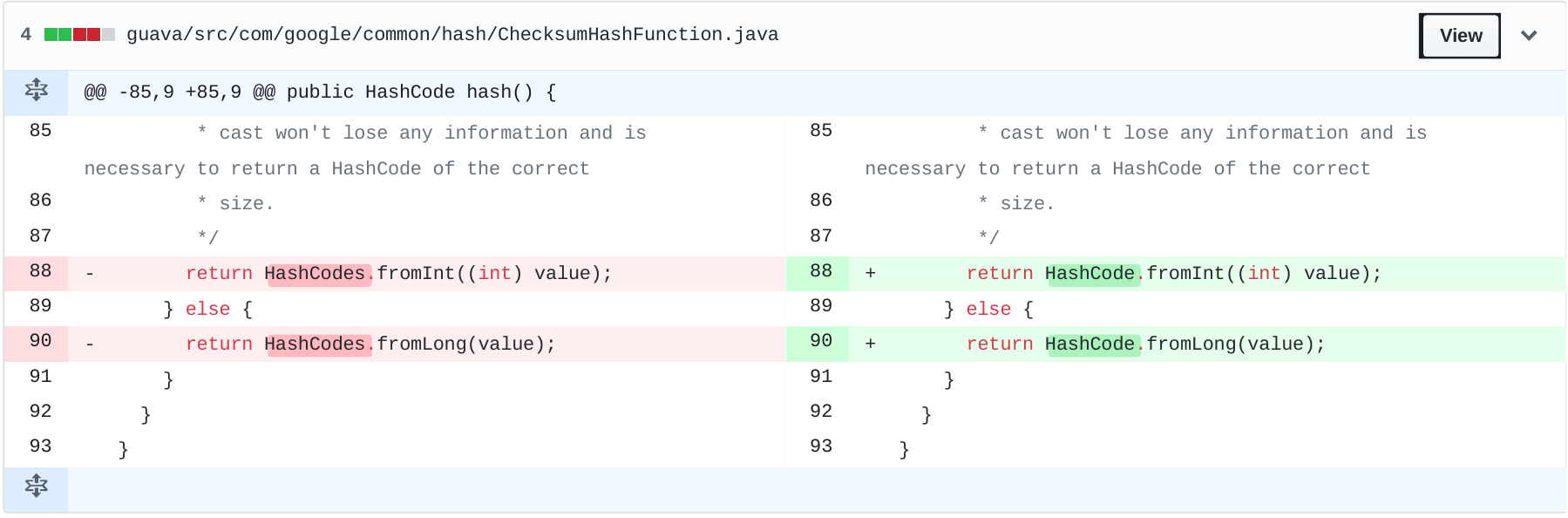}
	}\hfill
	\caption{We show two examples of API breaking changes that were not mapped to detected refactoring operations (i.e., \texttt{non-Ref}). We conjecture that these changes are (a)  in response to a complex defect in the code and (b) consist of a complex refactoring that is not captured by the automated approach.}
\label{fig:RQ3ex}	
	\centering
\end{figure}
 
From our manual analysis and similar to a study by Murphy-Hill et al. \cite{Murphy-Hill2009}, we find that not all API changes appear in the API change logs.
Figure \ref{fig:RQ3ex} does show two documented case examples of API breakages that are not mapped to a detected refactoring operation (i.e., \texttt{non-Ref}).
These examples provide evidence that these many client-used API class breakages are: (a)  motivated by a bug fix or new feature or (b) consists of a complex refactoring that is not captured by the automated approach.
In the first example (i.e., Figure \ref{fig:RQ3ex1}, we show an unavoidable API breaking change, especially if it is used to fix a complex defect such as a third party library.
This API breaking change was triggered in response to an error reported by a client user \textit{"JDK and Guava TypeVariable implementations are no longer compatible under 1.7.0 51-b13"} \footnote{issue at \url{https://github.com/google/guava/issues/1635} and fix at \url{https://goo.gl/bqDpxU}} It was widely reported to affect many client users of the library.
Developers found that a change in the standard Java library (JDK) causes guava to break API compatibility, as  prior guava version implemented an undocumented \textit{internal API} of the JDK (i.e., \texttt{Types.TypeVariable.newTypeVariable()})\footnote{A blogger discussions by users is at \url{https://goo.gl/8tcHfY}}. 
After much discussion among developers, the accepted API change was documented to \textit{`conditionally work only under the new JDK'}.

In the second example (i.e., Figure \ref{fig:RQ3ex2}), we acknowledge cases where the automated approach is unable to detect more complex refactoring operations.
Soares et al. \cite{Soares2013} showed that Ref-Finder is unable to correctly detect all types of refactoring operations, which is a validity threat and is discussed in detail (See Sections \ref{sec:challenge} and Section \ref{sec:threats}).
Moreover, this change is listed as a submitted enhancement issue\footnote{url{https://github.com/google/guava/issues/1495}} related to \textit{`Move HashCodes static methods to HashCode'} and involves 17 changed files (261 added and 219 deleted lines of code)\footnote{the code change is at \url{https://goo.gl/JHVi5J}}.  

%These results confirms that Ref-Finder may not be able to detect the more complex refactoring activities.

\begin{hassanbox}
	\label{sec:RQ3Sum}
Findings indicate that many client-used API breakages are likely to be motivated by other maintenance issues (i.e., bug fixes and new features) and involve more complex refactoring operations.
\end{hassanbox}

\begin{landscape}
	\begin{table*}
		\centering 
		\fontsize{10}{10}\selectfont
		\tabcolsep=0.1cm
		\caption{Classification of \RO~for API classes with \textit{presver} ratio. Note that one class may be classified under several refactoring types. Note (--) represents no matches. We also show the total of all breakages (cu. + ncu.) and use to colors to highlight when
			\colorbox{green!25}{prsv} = low and \colorbox{red!25}{prsv} = high.}
		\label{tab:refactoringTypes} 
		\begin{tabular}{@{}lccrccccrccccrccr} 
		\toprule 
			Classification of \RO &  \multicolumn{5}{c}{\textsc{guava}} & \multicolumn{5}{c}{\textsc{httpclient}} & \multicolumn{5}{c}{\textsc{xerces}}\\ 
			\cmidrule(r){2-6} \cmidrule(r){7-11} \cmidrule(r){12-16}
			
			&  \multicolumn{3}{c}{breaking} & \multicolumn{2}{c}{non breaking}  &  \multicolumn{3}{c}{breaking} & \multicolumn{2}{c}{non breaking}       &  \multicolumn{3}{c}{breaking} & \multicolumn{2}{c}{non breaking}    \\
			\cmidrule(r){2-4} \cmidrule(r){5-6} \cmidrule(r){7-9} \cmidrule(r){10-11} \cmidrule(r){12-14} \cmidrule(r){15-16}
			
			&  \multirow{2}{*}{\rotatebox[origin=c]{90}{\activeAPIAB}} & \multirow{2}{*}{\rotatebox[origin=c]{90}{\inactiveAPIAB}} &\multirow{2}{*}{\rotatebox[origin=c]{90}{prsv}} 
			&\multirow{2}{*}{\rotatebox[origin=c]{90}{\activeAPIAB}} & \multirow{2}{*}{\rotatebox[origin=c]{90}{\inactiveAPIAB}} &
			\multirow{2}{*}{\rotatebox[origin=c]{90}{\activeAPIAB}}& \multirow{2}{*}{\rotatebox[origin=c]{90}{\inactiveAPIAB}}&\multirow{2}{*}{\rotatebox[origin=c]{90}{prsv}} 
			&\multirow{2}{*}{\rotatebox[origin=c]{90}{\activeAPIAB}} & \multirow{2}{*}{\rotatebox[origin=c]{90}{\inactiveAPIAB}}& 
			\multirow{2}{*}{\rotatebox[origin=c]{90}{\activeAPIAB}} & \multirow{2}{*}{\rotatebox[origin=c]{90}{\inactiveAPIAB}}&\multirow{2}{*}{\rotatebox[origin=c]{90}{prsv}} 
			&\multirow{2}{*}{\rotatebox[origin=c]{90}{\activeAPIAB}} & \multirow{2}{*}{\rotatebox[origin=c]{90}{\inactiveAPIAB}}\\ 
%			&API&API& & API &API & API& API&&API&API&API&API&&API&API \\ 
			 \\
			\\
			\\
			\\
			\\ \hline
			change\_parameter & $28$ & $25$ &\cellcolor{red!25} (53) $1.12$ & $20$ & $4$ & $1$ & $10$ & \cellcolor{green!25}(11) $0.10$ & %$1$ & $-$ & $-$ & $-$ & - & - & $-$ & - & - & - & $-$ & - & - & - & - & $1$
			$-$& $11$ & $56$ & $217$ &\cellcolor{green!25} (273) $0.26$ & $-$ & $19$ \\ 
			cdcf* & $-$ & $-$ & $ $ & $-$ & $-$ & $-$ & $1$ & $ $ & $4$ & $1$ 
			%& $-$ & $-$ & - & - & $-$ & - & - & - & $1$ & - & - & - & - & $-$ & $-$ 
			& $44$ & $74$ & \cellcolor{green!25}(118) $0.59 $ & $8$ & $48$ \\ 
			extract\_method & $8$ & $32$ & \cellcolor{green!25}(40) $0.25$  & $3$ & $-$ & $-$ & $1$ & $ $ & $-$ & $2$ %& $-$ & $-$ & - & - & $-$ & - & - & - & $-$ & - & - & - & - & $-$ & $-$
			& $9$ & $61$ & \cellcolor{green!25} (70) $0.15$ & $5$ & $47$ \\ 
			extract\_subclass & $3$ & $-$ & $ $ & $-$ & $-$ & $-$ & $-$ & $ $ & $-$ & $-$
			%& $-$ & $-$ & - & - & $-$ & - & - & - & $-$ & - & - & - & - & $-$ & $-$
			& $-$ & $1$ & $ $ & $-$ & $-$ \\ 
			extract\_superclass & $-$ & $1$ & $ $ & $-$ & $-$ & $-$ & $-$ & $ $ & $-$ & $-$
			% & $-$ & $-$ & $-$ & - & - & $-$ & - & - & - & $-$ & - & - & - & - & $-$ & $-$ & 
			&$-$ & $2$ & $ $ & $-$ & $-$ \\ 
			inline\_method & $6$ & $3$ &\cellcolor{red!25} (9) $2.00$   & $-$ & $2$ & $1$ & $-$ & $ $ & $-$ & $-$
			% & $-$ & $-$ & - & - & $-$ & - & - & - & $-$ & - & - & - & - & $-$ & $-$ 
			& $7$ & $23$ &\cellcolor{green!25}  (30)  $0.30$& $1$ & $1$ \\ 
			inline\_temp & $4$ & $4$ &\cellcolor{gray!25}  (8) $1.00$  & $6$ & $1$ & $2$ & $-$ & $ $ & $-$ & $-$
			% & $-$ & $-$ & - & - & $-$ & - & - & - & $-$ & - & - & - & - & $-$ & $-$ 
			& $8$ & $20$ & \cellcolor{green!25}(28) $0.40$  & $-$ & $7$ \\ 
			introduce\_explaining\_variable & $-$ & $10$ & $ $ & $6$ & $2$ & $-$ & $-$ & $ $ & $-$ & 
			%$-$ & $-$ & $-$ & - & - & $-$ & - & - & - & $-$ & - & - & - & - & $1$ & $-$
			& $8$ & $33$ & \cellcolor{green!25} (41) $0.24$ & $2$ & $21$ \\ 
			introduce\_null\_object & $-$ & $-$ & $ $ & $1$ & $2$ & $-$ & $-$ & $ $ & $-$ & $-$
			% & $-$ & $-$ & - & - & $-$ & - & - & - & $-$ & - & - & - & - & $-$ & $-$ 
			& $-$ & $-$ & $ $ & $-$ & $-$ \\ 
			move\_field & $10$ & $66$ & \cellcolor{green!25}(76) $0.15$  & $9$ & $7$ & $-$ & $4$ & $ $ & $-$ & $-$
			% & $-$ & $-$ & - & - & $-$ & - & - & - & $-$ & - & - & - & - & $-$ & $-$
			& $15$ & $255$ & \cellcolor{green!25}(270) $0.06$  & $7$ & $5$ \\ 
			move\_method & $15$ & $19$ & \cellcolor{green!25}(34) $0.08$ & $8$ & $14$ & $3$ & $6$ &\cellcolor{green!25}(9)  $0.50$ & $-$ & $7$ 
			%& $-$ & $-$ & - & - & $8$ & - & - & - & $-$ & - & - & - & - & $-$ & $-$
			& $12$ & $256$ &\cellcolor{green!25}  (268) $0.05$ & $16$ & $12$ \\ 
			pull\_up\_constructor\_body & $-$ & $-$ & $ $ & $1$ & $-$ & $-$ & $-$ & $ $ & $-$ & $-$
			% & $-$ & $-$ & - & - & $-$ & - & - & - & $-$ & - & - & - & - & $-$ & $-$ & $-$ 
			& $-$ & $-$ & $ $ & $-$& $-$  \\ 
			pull\_up\_field & $-$ & $3$ & $ $ & $-$ & $3$ & $-$ & $-$ & $ $ & $-$ & $-$
			% & $-$ & $-$ & - & - & $-$ & - & - & - & $-$ & - & - & - & - & $-$ & $-$ & $-$ 
			& $-$ & $5$ & $ $ & $-$ & $-$ \\ 
			push\_down\_field & $-$ & $2$ & $ $ & $-$ & $-$ & $-$ & $-$ & $ $ & $-$ & $-$
			% & $-$ & $-$ & - & - & $-$ & - & - & - & $-$ & - & - & - & - & $-$ & $-$ 
			& $-$ & $41$ & $ $ & $-$ & $-$ \\ 
			ratp* & $-$ & $2$ & $ $ & $-$ & $3$ & $2$ & $-$ & $ $ & $2$ & $5$ 
			%& $-$ & $-$ & - & - & $-$ & - & - & - & $-$ & - & - & - & - & $-$ & $-$ 
			& $2$ & $14$ &\cellcolor{green!25}(16) $0.14$ & $-$ & $9$ \\ 
			remove\_control\_flag & $5$ & $1$ & \cellcolor{red!25}(5) $5.00$  & $1$ & $12$ & $2$ & $1$ & \cellcolor{red!25}(3) $2.00$  & $-$ & $5$ 
			%$1$ & $1$ & 1 & - & $-$ & - & - & - & $-$ & - & - & - & - & $-$ & $-$ 
			& $1$ & $11$ & \cellcolor{green!25} (12) $0.09$ & $1$ & $6$ \\ 
			remove\_middle\_man & $1$ & $-$ & $ $ & $-$ & $-$ & $-$ & $-$ & $ $ & $-$ & $-$ %& $-$ & $-$ & - & - & $-$ & - & - & - & $-$ & - & - & - & - & $-$ & $-$ 
			& $-$ & $-$ & $ $ & $-$ & $-$ \\ 
			remove\_parameter &$30$ & $20$ & \cellcolor{red!25}(50) $1.50$  & $16$ & $2$ & $1$ & $8$ & \cellcolor{green!25}(9) $0.12$  & $1$ & $-$ 
			%& $-$ & $-$ & - & - & $-$ & - & - & - & $-$ & - & - & - & - & $1$ & $11$ 
			& $45$ & $170$ &\cellcolor{green!25}(62) $0.26$  & $-$ & $18$ \\ 
			rename\_method & $28$ & $54$ &\cellcolor{green!25}(82) $0.52$  & $1$ & $4$ & $-$ & $-$ & $ $ & $2$ & $-$ 
			%& $-$ & $-$ & - & - & $-$ & - & - & - & $-$ & - & - & - & - & $-$ & $-$ 
			& $146$ & $217$ & \cellcolor{green!25}(363) $0.67$ & $1$ & $19$ \\ 
			rcwfm* & $2$ & $2$ & \cellcolor{gray!25}(4) $1.00 $ & $12$ & $7$ & $-$ & $-$ & $ $ & $-$ & $-$ 
			%& $-$ & $-$ & - & - & $-$ & - & - & - & $-$ & - & - & - & - & $-$ & $-$ 
			& $-$ & $-$ & $ $ & $-$ & $-$ \\ 
			replace\_data\_with\_object & $2$ & $-$ & $ $ & $-$ & $-$ & $-$ & $-$ & $ $ & $1$ & $-$
			% & $-$ & $-$ & - & - & $-$ & - & - & - & $-$ & - & - & - & - & $-$ & $-$ 
			& $4$ & $11$ & \cellcolor{green!25}(15) $0.36$ & $-$ & $2$ \\ 
			replace\_exception\_with\_test & $-$ & $9$ & $ $ & $-$ & $-$ & $-$ & $1$ & $ $ & $-$ & $-$
			% & $-$ & $-$ & - & - & $-$ & - & - & - & $-$ & - & - & - & - & $-$ & $-$
			& $-$ & $3$ & $ $ & $-$ & $4$ \\ 
			rmnwc* & $16$ & $34$ &\cellcolor{green!25}(50) $0.47$  & $8$ & $20$ & $-$ & $8$ & $ $ & $-$ & $17$ 
			%& $-$ & $-$ & - & - & $1$ & - & - & - & $2$ & - & - & - & - & $2$ & $-$
			& $49$ & $165$ & \cellcolor{green!25}(214) $0.30$ & $9$ & $117$ \\ 
			rmwmo* & $4$ & $-$ & $ $ & $2$ & $-$ & $-$ & $-$ & $ $ & $1$ & $-$
			% & $-$ & $-$ & - & - & $-$ & - & - & - & $-$ & - & - & - & - & $-$ & $-$
			& $4$ & $40$ & \cellcolor{green!25} (44) $0.10$& $1$ & $5$ \\ 
			rncgc* & $-$ & $-$ & $ $ & $-$ & $1$ & $-$ & $-$ & $ $ & $-$ & $-$
			% & $-$ & $-$ & - & - & $-$ & - & - & - & $-$ & - & - & - & - & $-$ & $-$ 
			& $9$ & $27$ & \cellcolor{green!25}(36) $0.33$ & $-$ & $14$ \\ 
			replace\_temp\_with\_query & $-$ & $6$ & $ $ & $1$ & $-$ & $-$ & $-$ & $ $ & $-$ & $-$ 
			%& $-$ & $-$ & - & - & $-$ & - & - & - & $-$ & - & - & - & - & $-$ & $-$ 
			& $-$ & $-$ & $ $ & $-$ & $1$ \\ 
			pull\_up\_method & $-$ & $-$ & $ $ & $-$ & $-$ & $2$ & $1$ & \cellcolor{red!25} (3) $2.00$  & $-$ & $-$ 
			%& $-$ & $-$ & - & - & $-$ & - & - & - & $-$ & - & - & - & - & $-$ & $-$ 
			& $5$ & $2$ &\cellcolor{red!25}(7) $2.50$  & $-$ & $-$ \\ 
			extract\_interface$^a$ & $-$ & $-$ & $ $ & $-$ & $-$ & $-$ & $-$ & $ $ & $-$ & $-$ %& $-$ & $-$ & - & - & $-$ & - & - & - & $-$ & - & - & - & - & $-$ & $-$ 
			& $-$ & $1$ & $ $ & $9$ & $-$ \\ 
			\hline \\
			Median ($\bar{x}$) & $4$ & $4$ & \cellcolor{gray!25}$1.00$ & $3$ & $3$ & $2$ & $1$ & \cellcolor{green!25}$0.50$ & $1$ & $5$ 
			& $9$ & $17$ & \cellcolor{green!25}$0.33$ & $5$ & $11$ \\ 
			Mean ($\mu$) & $8.4$ & $14.78$ & \cellcolor{red!25} $1.19$ & $5.14$ & $4.4$ & $1.75$ & $3.2$ & \cellcolor{green!25}$0.95$ & $1.71$ & $6.17$ 
			& $24.9$ & $62.70$ & \cellcolor{green!25}$0.46$ & $5.45$ & $19.72$ \\ 
			\\\hline\\ 
		\end{tabular} 
		\\
		\footnotesize{$^*$Note types abbreviations - cdcf = consolidate\_duplicate\_cond\_fragment, rcwfm = replace\_constructor\_with\_factory\_method, ratp = remove\_assignment\_to\_parameters, rmnwc = replace\_magic\_number\_with\_constant, rmwmo = replace\_method\_with\_method\_object, rncgc = replace\_nested\_cond\_guard\_clause \\[-1.8ex]}
	\end{table*} 
\end{landscape}

%\paragraph{\underline{Findings:}}

\subsection{Findings for RQ4}

\sloppypar{
Table \ref{tab:refactoringTypes} shows a classification of the applied \RO~for the three libraries \textsc{guava}, \textsc{xerces}, and \textsc{httpclient} in our collected dataset.
As seen in the table, \textsc{guava} developers applied the \texttt{change\_parameter} \RO~ 28 times to breaking \activeAPI. Developers subsequently applied the same \RO~and broke 25 \inactiveAPI~during the library evolution of \textsc{guava}.
We find that the \textsc{guava} and \textsc{xerces} libraries tend to refactor and break their versions during evolution than \textsc{httpclient}. 
Our results align with the findings of Cossette \textit{et al.} \cite{Cossette2012} on API transformations, where they also used the same libraries in their experiments. 
Results in Table \ref{tab:refactoringTypes} show that developers apply specific \RO~more frequently when evolving their libraries.
For instance, \RO~such as \texttt{move\_method} (\textsc{guava}- 34 \RO), \texttt{change\_parameter} (\textsc{httpclient}- 11 \RO), and \texttt{rename\_method} (\textsc{xerces}- 363 \RO) were the most frequently applied that cause API breakages.
 For \activeAPI, \texttt{remove\_parameter} (\textsc{guava}- 30 \RO), \texttt{move\_method} (\textsc{httpclient}- 3 \RO) and \texttt{rename\_method} (\textsc{xerces}- 146 \RO) are reported as most frequent. 
Notably, \texttt{move\_method} (\textsc{guava}- 190 \RO, \textsc{xerces}- 256 \RO), \texttt{remove\_parameter} (\textsc{httpclient}- 8 \RO) were applied to \inactiveAPI. }
%\begin{hassanbox}
%\label{sec:RQ3SumA}
%Refactoring operations associated with parameters and methods are more likely to break \activeAPI.
%\end{hassanbox}

Table \ref{tab:refactoringTypes} also reports the \textit{prsv} ratio for each library. 
This metric measures the degree of likelihood to which library developers apply certain \RO~to \activeAPI~compared to \inactiveAPI~(\ie~preserving  \activeAPI). 
We use color to highlight the \textit{prsv} scores.
Green highlights in the table represents a low preservation of \activeAPI, while the red highlights indicates a high ratio of \RO~in \inactiveAPI.
For example, the library developers of both \textsc{guava} (\textit{prsv}= 0.08) and \textsc{xerces} (\textit{prsv} = 0.05) tend to apply less \texttt{move\_method} refactoring operations to \activeAPI. 
Our results shows library maintainers are less likely to refactor (using the more frequent \RO) \activeAPI~than \inactiveAPI.
For example, 5 out of 10 \RO~in \textsc{guava},  3 out of 5 \RO~types in \textsc{httpclient}, and 16 out of 17 \RO~types in \textsc{xerces} are less likely applied to \activeAPI. 
We find that many high \textit{prsv} ratios (depicted by red in the table) where by the rarely applied \RO~types (e.g., \texttt{remove\_control\_flag} (\textsc{guava}- 5 \RO, \textsc{httpclient}- 2 \RO) and \texttt{pull\_up\_method} (\textsc{httpclient}- 3  \RO, \textsc{xerces}- 7 \RO).

%\vspace{-2mm}
\begin{hassanbox}
\label{sec:RQ3Summary}
Findings show that library maintainers were more likely to refactor \inactiveAPI~compared to \activeAPI. 
\end{hassanbox}

\section{Discussion}
\label{sec:discussion}
In this section, we first discuss the implications of results and then compare with related work.
We then discuss some challenges of our approach and finally present threats to the validity of our study.

\subsection{Implications}
%We discuss the implications from two viewpoints of both client and library developers as well as the researcher. 
Our results indicate that when evolving libraries, out of all code changes applied, maintainers are less likely to apply incompatible code changes to external API classes compared to the other classes during the library evolution.
This implies that library developers may understand the efforts by clients needed to update their libraries.
Complementary to this finding, Bloch mentions the growing awareness of library maintainers to APIs \cite{Bloch:2006}. 
This is also reinforced by Seo \textit{et al.} \cite{Seo:2014} where they found that there are many cases where API breakage changes are only applied when unavoidable (\ie~in response to either vulnerabilities or needed bug fixes etc...).
There are benefits to this awareness of client-used API breakages. 
In particular, the evolution of APIs encourages trust and reduce the latency of adoption by client projects, which is currently being experienced as a problem by many OSS clients \cite{KulaSANER2014}.
Larman \cite{LarmanPV} introduced a notion of the Protected Variation (PV) pattern: identify points of predicted variation and create a stable interface around them. 
This PV pattern could explain how contemporary developers build and evolve libraries in relation to client-used APIs.

\subsection{Comparison to Literature}
It is important to understand that our work cannot be simply compared at face-value to prior studies.
As outlined in Section \ref{sec:motiv} there are obvious differences with our approach, compared to the studies of Dig and Johnson \cite{Dig2006} and Cossette \textit{et al} \cite{Cossette2012}.
Dig and Johnson used the change logs as heuristic to locate all API changes, and other considered public entities are APIs. 
In this study, we detect syntactic changes in classes to infer changes and determine client usage to identify if the change has an effect to its users. 
As a result, our approach is unable to detect behavioral API breakages. 
Dig and Johnson's study included behavioral breakages, which we do not consider due to the limitations of our approach. 
Our definition of an API does differ from prior work.
Dig and Johnson considered all public entities. 
Our work are more similar to the work of Cossette \textit{et al.}, in which we include the detection of protected entities.
In this work, we go further and use client usage to focus on API breakages the more popular APIs.

The usage of tools revealed more API breakages, some of which were not reported in the API change logs, which was also consistent with the findings of Murphy-Hill \textit{et al} \cite{Murphy-Hill2009}.
%In recent times, many library developers utilizing these API breakage detection tools to determine and report all API changes.
These undocumented API changes could also explain the disparity in results between manual (\ie~Dig and Johnson study) and machinery refactoring detection.
For mechanical refactoring detection, since \texttt{Ref-Finder} is template-based refactoring reconstructing approach, we were only able to identify 23 out of 70 of Fowler's catalog. 
In fact, Cossette \textit{et al.} \cite{Cossette2012} also believed that tools would miss some behavioral refactoring, saying that they \textit{`...do not believe that some changes would be easily handled by mechanical transformation tools; instead the API maintainer, or the client developer would need to craft some minimal specification that would describe how to remap classes to accommodate these breaking changes.'}
Another difference in our method that may have influenced results, is where we analyze API changes between consecutive versions, while prior work analyzed versions that were not consecutive.

\begin{figure}[t]
	\centering
	\includegraphics[width=1\textwidth]{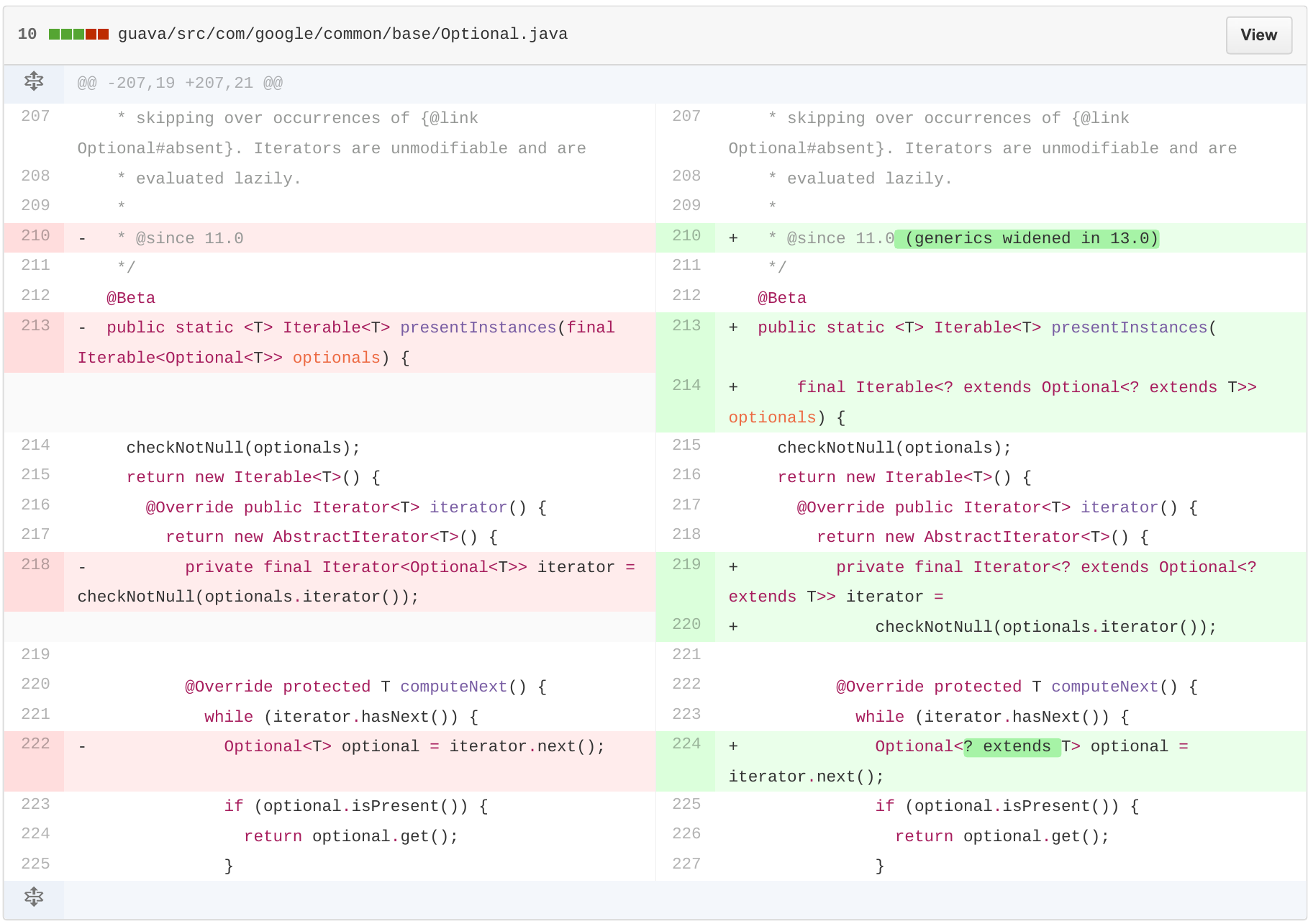}
	\caption{Example of a misidentified detected refactoring-related API breaking change (\texttt{Ref} $\cap$ \texttt{breaking} class).\texttt{Ref-Finder} detects this code change between \textsc{Guava} version 12 and 13 as an \texttt{add\_parameter} \RO, while the API breaking tool reports it as \texttt{ binary incompatible modified method}.}
	\label{fig:exampleRefAPIBreak} 
\end{figure}

\subsection {Challenges of the Automated Approach}
\label{sec:challenge}
%Two of the key findings learned from our study is that there exists complex refactoring operations not captured by an automated tool such as Ref-Finder and that not all API changes are documented by developers.
%Improvement of these tools would make breakthroughs in understanding how developers handle and treat API breakages.
Key threats to the automated approach accuracy is when: (i) refactorings are missed by our approach (i.e., Ref-Finder \cite{Soares2013}), (ii) developers may not report all API changes \cite{Murphy-Hill2009} and  (iii) misidentification of breaking APIs is reported but it did not cause a breakage.
Figure \ref{fig:exampleRefAPIBreak} presents an example of a misidentification reported by our automated approach \footnote{ commit can be found at \url{goo.gl/CwXoBj} and API change at \url{https://goo.gl/VPPTIX}} . 
In this example, \texttt{Ref-Finder} detects this change as an \texttt{add\_parameter} \RO, that is also API breaking since it has a change in the method signature. 
However, according to Java documentation, the superclass \texttt{extends} indicates that the change of this special \textit{`type parameter of the class does not, in itself, have any implications for binary compatibility'}. 
We believe these limitations will encourage researchers to further investigate and help us understand how developers evolve their libraries, especially in regards to avoidable API breakages.

%Secondly, out of refactoring operation 52 types defined by \texttt{Ref-Finder}only able to detect up to 29 types of the refactorings.
%Also, our results show that this only accounts for up to 37\% of all API breakages. 
%As found in RQ3, we were able to map up to 82\% as either bug fixes or issues, hence, one possibility is that there are some refactorings that are not yet detectable by our current state--of--the--art toolings.

%The most related work of APIs and refactoring is the study that was performed by Dig and Johnson . In their work, they manually inspected library release notes for documented API changes between two versions of a library framework. We consider the two works different because: (i) our automated approach also detects undocumented API breakages and the ref-finder tool may provide different results compared to manual analysis and (ii) we distinguish between external and internal APIs. Yet, our results are consistent with \cite{Cossette2012,Kapur2010} and draw similar conclusions of developers awareness by Bloch \cite{Bloch:2006} and Seo \textit{et al.} \cite{Seo:2014}.
 
\subsection{Threats to Validity}
\label{sec:threats}
\textbf{Internal Threats:}
The most significant internal threat is correctness of the automated tools, especially \texttt{Ref-Finder}.
To mitigate this and as a sanity check, we randomly inspected a small sample of the results for validation. 
%Concretely, for each library version, we semi-automatically inspected to eliminate any false positives. 
Mentioned earlier in the paper, an example of a false positive was when a unchanged file was reported to have a refactoring identified. 
In the end, we understand that recall is not as obvious to investigate as ground truth is unknown.
\texttt{Ref-Finder} is the current state--of--the--art and actively used in research.

Another minor threat to our approach is that API breakages false positives caused by the class-level granularity of analysis. Theoretically, an external API class that has a breakage related to a private entity could be a false positive. However, even with this assumption in mind, our analysis may be underestimations.
It is true that the accuracy of the saturation point is fairly dependent on the sample size. We believe our sample clients are sufficient to at least identify the most popular APIs that reside in the \activeAPI.
Sometimes variations between the refactored classpath (originating from source code) and API breakages class path (originating from binary code) may cause a miss-match. 
To overcome this, we manually validated the consistency of file paths to ensure consistency and completeness.  
Correct ordering of consecutive library releases is another minor threat. 
We therefore consider Maven \cite{MavenCentralURL} as the ground truth to base our chronological ordering of the released versions of a library. 
Some of our conclusions are based on the statistical analysis. 
We believe that due to outliers and nature of the data collected, non parametric statistical tests were deemed appropriate. 
%Understanding the nature of the rest of API breakages was in the scope of this study. 
%However, understanding these other API breakages could lead to other categories of changes other than the current set of known refactorings. This could be a potential avenue for future work.
%To mitigate bias, we present all datasets in boxplots for visual analysis.

\textbf{External Threats:}
As an external threat, we understand that our collected clients and the six selected OSS libraries are not necessary complete representations of the real world.
However,  we believe that the diverse nature (such as size, domain, team) of the six libraries is enough to assume generalization. 
Although our approximations of external APIs can only be justified through documentation and developers, we believe our method provides sufficient confidence of external client coverage.
Another important threat is selection of the more popular libraries. As a results, our findings may not be applicable for less popular libraries. In this study, we consider that both library developers and users are more concerned with popular APIs, as they tend to reach a larger client user-base. Moreover, the same libraries that we study have been used in prior studies by researchers. As future work, we plan to expand our study to investigate more frameworks and libraries.
Since our study is focused only on java libraries, we cannot make generalizations to other programming languages. 
We are confident that our research method is scalable and can be replicated with different sets of clients and subject libraries in other languages.

\section{Related Work}\label{sec:related}
In this section, we introduce literature related to API usage, library migration support and library evolution.

\textbf{API usage}.
There has been different work that have collected clients API usage. For example, work such as De Roover \textit{et al.} \cite{DeRoover2013} exploit API usage to understand popularity and usage patterns of clients. The data collected is visualized to further explore to provide program comprehension as well as identify patterns in the code. 
Another set of research use the API usage as a measure of stability or popularity \cite{McDonnell2013,Mileva:2009}. Our previous work \cite{2014VISSOFTKula}, among work leveraged popularity to recommend when libraries are deemed safe to use by the masses. 
Other related work that studied the impact of API evolution on their clients on online forums such as Stack Overflow \cite{DBLP:conf/iwpc/VasquezBPOP14} and the  Android App \cite{DBLP:journals/tse/BavotaVBPOP15}, Pharo \cite{DBLP:conf/icsm/HoraRAEDV15} and Smalltalk ecosystems \cite{DBLP:conf/sigsoft/RobbesLR12}.

%Similar to these related work, we leverage API usage to classify `active' APIs.

%Similarity, we use API usage as a measure of stability 

%Cox \textit{et al.} \cite{Cox:2015} also studied dependency usage of clients. Their work uses defined dependencies and not compiler loaded dependencies of classes. Also the analysis is not at the class-level. Also, all public entities are considered APIs.

%\textbf{Library API Documentation} It is known that `code is king', as online documentation can become quickly outdated. Thus studying how APIs are used and patterns serve as a learning resource. Work by Sub...

\textbf{Library Migration Support}.
Much work has been in transformation of the client code to support migration of library API changes. Work by Chow and Notkin \cite{Chow:1996} and Balaban \textit{et al.} \cite{Balaban:2005} use a change specification language. 
There is work that provides the client with automatic tool support to accommodate changes made the APIs of a library. For instance, SemDiff \cite{Dagenais:2009} recommends replacements for framework methods that were accessed by clients. Other similar tools were proposed by Xing and Stroulia \cite{Xing2007} and Schafer \textit{et al.} \cite{Schafer:2008}. Other work on reuse support is through code analysis. This area of work considers code clone detection techniques \cite{KamiyaTSE2002} to support which library version is most appropriate candidate for migration. Godfrey \textit{et al.} \cite{Godfrey2005} proposed origin analysis to recover context of code changes. Our previous work \cite{Kawamitsu2014} tracked how code is reused cross-projects.
Related works \cite{Kapur2010} focused on support for clients migrating to a newer library version.
Likewise, other works  \cite{McDonnell2013,Malpohl:2000,Mezini:1997,Steyaert:1996} studied how library maintainers balance API compatibility with an evolving library. 

\textbf{Library Evolution}.
There is similar work with respect to library maintenance and evolution. Cossette \textit{et al.} \cite{Cossette2012} manually illustrated the complexities of library changes and transformations. Other work such as Kim \textit{et al.} \cite{Kim2011} studied the role of refactoring during software evolution. 
Recently, there has been large-scale empirical studies conducted on library migrations and evolution. Empirical studies by Raemakers \textit{et al.} \cite{Raemaekers2014,RaemaekersICSM}, Jezek \textit{et al.} \cite{Jezek2015} and Joel \textit{et al.} \cite{Cox:2015} studied in-depth how libraries that reside in the Maven Central super-repository evolve and break APIs. 

\section{Conclusions and Future Work}
\label{sec:conclusion}
%context
Refactorings is a key maintainability practice, even for library maintainers.
When evolving code, we find that library developers are less likely to break APIs. 
However, we find that many of these API breaking changes relate to bug fixes and new features, with only up to 37\% of client-used API breakages related to refactoring operations. 
The study finds that there are still challenges to improving our tools. 
The study also reveals challenges faced by the tools. As future work, we envision that this study encourages more research into automated refactoring detection techniques to advance our understanding of refactoring activities on API breakages.

\section{Acknowledgments}
This work is supported by JSPS KANENHI (Grant Numbers JP25220003 and JP26280021) and the ``Osaka University Program for Promoting International Joint Research.".
%Ali Ouni is supported by the `Research Start-up (2) 2016 Grant G00002211' funded by UAE University.

\bibliographystyle{abbrv}
\bibliography{sigproc}

\end{document}